\documentclass[preprint,nofootinbib,aps,superscriptaddress]{revtex4}
\usepackage{amsmath,amssymb}
\usepackage{mathtools}
\usepackage[usenames, dvipsnames]{xcolor}
\usepackage{braket}
\usepackage{graphicx}
\usepackage{slashed}
\usepackage{tensor,siunitx}
\usepackage{changes}
\usepackage{subfigure}
\usepackage{multirow}
\usepackage{float}
\usepackage{threeparttable}
\usepackage[colorlinks=true,linkcolor=red,citecolor=blue]{hyperref}

\usepackage{comment}
\usepackage[normalem]{ulem}

\newcommand{\be}{\begin{equation}}
\newcommand{\ee}{\end{equation}}

\newcommand{\bea}{\begin{eqnarray}}
\newcommand{\eea}{\end{eqnarray}}

\allowdisplaybreaks

\begin{document}

\title{\boldmath
Extended parity doublet model with a new transport code}

\author{Myungkuk Kim}
\email{myung.k.kim@pusan.ac.kr}
\affiliation{Department of Physics, Pusan National University, Busan 46241, Korea}

\author{Sangyong Jeon}
\affiliation{Department of Physics, McGill University, Montreal, Quebec, H3A2T8, Canada}

\author{Young-Min Kim}
\affiliation{Department of Physics, Ulsan National Institute of Science and Technology, Ulsan, 44919, Korea}

\author{Youngman Kim}
\affiliation{Rare Isotope Science Project, Institute for Basic Science, Daejeon 34047, Korea}

\author{Chang-Hwan Lee}
\email{clee@pusan.ac.kr}
\affiliation{Department of Physics, Pusan National University, Busan 46241, Korea}

\begin{abstract}
A new transport code ``DaeJeon Boltzmann-Uehling-Uhlenbeck (DJBUU)" had been developed
and enables to describe the dynamics of heavy-ion collisions in low-energy region.
To confirm the validity of the new code, we first calculate Au + Au collisions at $E_{\rm beam}=100$ and $400A$ MeV
and also perform the box calculation to check
the detail of collisions and Pauli blocking without mean-field potential as suggested by
the Transport Code Comparison Project.
After confirming the validity of new transport code, we study low-energy heavy-ion collisions with
an extended parity doublet model.
Since the distinctive feature of the parity doublet model 
is the existence of the chiral invariant mass that contributes to the nucleon mass,
we investigate how physical quantities depend on 
the chiral invariant mass in heavy ion collisions at low energies.
For this, we calculate physical quantities such as
the effective nucleon mass in central collisions
and transverse flow in semi-central collisions of Au + Au at 
$E_{\rm beam}=400A$ MeV with different values of the chiral invariant masses.
\end{abstract}

\pacs{
}

\date{\today}

\maketitle

\section{Introduction}


Understanding asymmetric nuclear matter is one of the key issues in contemporary nuclear
physics. The study of exotic
nuclei, compact starts, core-collapsed supernovae and many facets of the QCD phase diagram
all critically depend on such understanding.
Forthcoming facilities ~\cite{Motobayashi:2014uja} such as RAON, FRIB, FAIR, and RIKEN RIBF
will be creating highly asymmetric nuclear matter by colliding heavy
ions for the goal of understanding the neutron rich matter.

Heavy-ion collisions (HICs) offer a great opportunity to do
researches in a wide range of the densities, temperatures, and isospin asymmetries.
However, some important quantities in dense matter studies such as 
the nuclear symmetry energy and its slope parameter are
not directly accessible in such experiments.
An important way to extract such information from the HICs
is to use nuclear transport simulations to test out various scenarios.
The purpose of this work is to study asymmetric nuclear matter using 
the BUU (Boltzmann-Uehling-Uhlenbeck) approach.

Transport theories have been applied to heavy-ion collision simulations
since 1980's~\cite{Bertsch:1984gb, Kruse:1985pg, Aichelin:1986zz}.
Currently, two types of transport approaches are in wide use.
One is the the BUU approach which evolves the one-particle phase
density by propagating test particles in the mean-fields between collisions.
The other is the QMD (Quantum Molecular Dynamics) approach which attempts to evolve particles
according to the given many-body Hamiltonian.
In order to understand and reduce the uncertainty between different codes,
a few transport code comparison projects have been carried out over the years
~\cite{Kolomeitsev:2004np, Xu:2016lue, Zhang:2017esm, Ono:2019ndq}.
In this work, we will first compare our results to those 
in Ref.~\cite{Xu:2016lue} and Ref.~\cite{Zhang:2017esm}
in Section~\ref{sec:DJBUU_comparison}
to ensure that our BUU model
is performing within the established norm before applying it to the
extended parity doublet model in Section~\ref{sec:pdm}.

There are several existing BUU (Boltzmann-Uehling-Uhlenbeck) codes 
developed for heavy-ion collisions, 
such as GIBUU~\cite{Gaitanos:2010fd, Larionov:2011fs, Buss:2011mx}, 
IBUU~\cite{Li:2008gp, Li:1997rc, Li:1997px, Chen:2013uua}, and
RBUU~\cite{Fuchs:1995fa, Gaitanos:2003zg, Ferini:2006je}.  
In this paper, we use the newly developed DJBUU (DaeJeon Boltzmann-Uehling-Uhlenbeck) 
code.
This code is optimized, for the moment, to HICs up to a few hundreds $A$ MeV.


As an application of DJBUU in heavy ion collisions, we study
the extended parity doublet model in this work.
The parity doublet model was formulated in Refs.~\cite{Detar:1988kn, Jido:2001nt}, and applied to the dense matter
in Refs.~\cite{Hatsuda:1988mv, Zschiesche:2006zj, Dexheimer:2007tn, Sasaki:2010bp, Gallas:2011qp, Steinheimer:2011ea, Benic:2015pia, 
Motohiro:2015taa, Takeda:2017mrm, Takeda:2018ldi, Marczenko:2018jui}.
As is well known, the mass of current quarks 
can explain only about 2\% of the
nucleon mass and the rest may be explained by other effects
such as the spontaneous chiral symmetry breaking.
In the parity doublet model, the nucleon mass has a contribution 
from the chiral invariant mass, apart from
the contribution from spontaneous chiral symmetry breaking.
At present, the origin of the chiral invariant mass is not well understood and its value is 
yet uncertain.

Since the chiral symmetry is expected to be partially restored in dense matter, 
change of the nucleon mass, caused by 
the reduction in the chiral condensate,
results in the change of observables in HICs.
Therefore, in order to constrain the value of the chiral invariant mass in the parity
doublet model,  
it is important to investigate the effect of the partial chiral symmetry restoration in low-energy heavy ion collisions. 
In Ref.~\cite{Zschiesche:2006zj}, the chiral invariant mass was estimated 
to be $m_0\sim 800$~MeV using the nuclear matter properties, especially incompressibility.
In an extended parity doublet model~\cite{Motohiro:2015taa},
the properties of nuclear matter were reproduced reasonably well with the chiral invariant
mass in the range from $500$ to $900$~MeV. 
In this work, we implement the extended parity doublet model ~\cite{Motohiro:2015taa, Shin:2018axs} 
in the DJBUU code and simulate heavy ion collisions with various values of the chiral invariant mass 
in an effort to better understand its value.

In Sec.~\ref{sec:DJBUU}, we introduce newly developed 
transport code DJBUU including basic equations and numerical schemes.
In Sec.~\ref{sec:DJBUU_comparison}, 
we compare our results of DJBUU in both HICs and box calculations 
with those of 
the Transport Code Comparison Project (TCCP)~\cite{Xu:2016lue}.
In Sec.~\ref{sec:pdm}, we summarize basic formalism of 
the extended parity doublet model implemented in the new transport code and
parameter sets extracted from nuclear structure calculation with the parity doublet model.
In Sec.~\ref{sec:result}, we present our results of the time evolution of 
mass splitting and anisotropic transverse flow with various values of the chiral invariant mass.
In Sec.~\ref{sec:conclusion}, final conclusion and discussion are summarized.

\section{DJBUU code description}
\label{sec:DJBUU}

\begin{table*}
\begin{center}
\begin{threeparttable}
\begin{tabular}{l c c c c c c c c c } 
\hline
\hline
parameter  & $~~f_\sigma~$(fm$^2)~~$ & $~~f_\omega~$(fm$^2)~~$ & $~~f_\rho~$(fm$^2)~~$ & $~~A~$(fm$^{-1})~~$ & $~~B~~$ & $~~m_N~~$ & $~~m_\sigma~~$ & $~~m_\omega~~$  & $~~m_\rho~~$  \\
& $10.33$ & $5.42$ & $0.95$ & $0.033$ & $-0.0048$ & $0.938$ & $0.5082$ & $0.783$ & $0.763$\\
\hline
\hline
\end{tabular}
\caption{Mean field parameter set and vacuum masses of all mesons in DJBUU taken from Ref.~\cite{Liu:2001iz}.
Coupling constants of mesons are defined as $f_i  \equiv (g_i^2 / m_i^2)$, $i=\sigma,\,\omega,\,\rho$ and $\sigma$ self-interaction terms
are $A \equiv a / g_\sigma^3$ and $B \equiv b/g_\sigma^4$. All the dimensions of masses are [GeV].}
\label{table:MF_set}
\end{threeparttable}
\end{center}
\end{table*}

In this section, we introduce the recently developed new transport code DJBUU.
The relativistic BUU equation with the mean field potential is given by

\be
\left[{p^\mu \partial_\mu^x - \left({p_\mu F^{\mu\nu} - m^*_i\partial^\nu_x m^*_i}\right) {\partial_\nu^p}} \right]  
{f_i({\bf x},{\bf p};t) \over E}
= {\cal C}^i_{\rm coll}\, ,
\label{eq:rbuu}
\ee
where $f_i({\bf x},{\bf p};t)$ is the phase space density of the hadron species $i$,
$F^{\mu \nu} \equiv \partial^\mu V^\nu - \partial^\nu V^\mu$ is the field strength tensor
associated with the vector meson mean-field $V^\mu$, and
$m^*_i$ is the effective mass of the $i$-th hadron species 
that includes the effect of the space-time dependent chiral condensate.
The superscript $x$ and $p$ on the partial derivatives indicate the spatial $(x)$ and the
momentum $(p)$ derivatives.
All possible collision processes 
including hadron $i$ and other hadron species $j$ are
described by collision term, ${\cal C}^i_{\rm coll}$.
For example, the elastic collision between
two baryon species
$i$ and $j$ 
is described by
\begin{widetext}
\begin{eqnarray}
C_{ij} 
& =&
{1\over 2}
\int {d^3p'_1\over (2\pi)^3 2E_{p'_1}}
\int {d^3p_2\over (2\pi)^3 2E_{p_2}}
\int {d^3p'_2\over (2\pi)^3 2E_{p'_2}}
\left| {\cal M}_{ij} \right|^2\,
(2\pi)^4\delta(p_1+p_2 - p'_1-p'_2)\,
\nonumber
\\
& & {} \times
\left[
f_i({\bf p}_1')f_j({\bf p}_2')\{1 - f_i({\bf p}_1)\}\{1 - f_j({\bf p}_2)\}
-f_i({\bf p}_1)f_j({\bf p}_2)\{1 - f_i({\bf p}_1')\}\{1 - f_j({\bf p}_2')\}
\right]\, .
\nonumber\\
\label{eq:collision}
\end{eqnarray}
\end{widetext}
where we suppressed the common $t$ and ${\bf x}$ dependence in the phase space densities for the
sake of brevity.
The first term in Eq.(\ref{eq:collision}) describes the collision process 
in which the energy level defined by the momentum ${\bf p}_1$ gains a particle,
and the second term in Eq.(\ref{eq:collision}) describes the collision process in which the 
energy level defined by the momentum ${\bf p}_1$ loses a particle. The $(1-f)$ factors
associated with the final state particles implement Pauli-blocking.
The scattering matrix element ${\cal M}_{ij}$ we use is the tree-level in-vacuum matrix
elements.

To solve for the phase space density, $f_i({\bf x}, {\bf p}; t)$, we use test particle method
which was firstly introduced to HIC simulations 
by Wong~\cite{Wong:1982zzb} in the early 1980s.
In this method, each physical particle is split into $N_{\rm test}$ test
particles. 
Hence, the phase space $\hat{f}_i$ and the cross-section $\hat{\sigma}_{ 1,2\rightarrow 1' ,2'...N'  }$
used in the simulation are scaled as
\bea
\hat{f}_i({\bf x},{\bf p};t) &=& f_i({\bf x}, {\bf p};t)  / N_{\rm test}\,, \\
\hat{\sigma}_{ 1,2\rightarrow 1' ,2'...N'  } &=& \sigma_{ 1,2\rightarrow 1' ,2'...N'  }/N_{\rm test} \,,
\eea
where $ f_i({\bf x}, {\bf p};t)$ and $\sigma_{ 1,2\rightarrow 1' ,2'...N'  }$ are
the physical phase space density and the cross-section, respectively.
In this work, we take 100 test particles for each nucleon ($N_{\rm test}$ = 100) and 
perform 10 independent simulations.
The simulated phase space density is represented by
\be
\hat{f}_i({\bf x},{\bf p};t) = {{(2\pi)^3} \over{N_{\rm test}}}\sum_{\alpha=1}^{N}\, 
g_x({\bf x} - {\bf x}_\alpha(t)) g_p({\bf p} - {\bf p}_\alpha(t))\, ,
\label{eq:hatf}
\ee
where $N$ is the total number of test particles and
${\bf x}_\alpha$ and ${\bf p}_\alpha$ are the coordinate and momentum of the $\alpha$-th test
particle, respectively. 
The functions
$g_x$ and $g_p$ are the profile functions
in the coordinate and momentum spaces.
In DJBUU, the following polynomial function is used for the profile 
instead of the often used Gaussian function: 
\begin{equation}
g({\bf u}) = g(u) = {\cal N}_{m,n} (1 - (u/a_{\rm cut})^m)^n
\ \ \ \hbox{for}\  0 < u/a_{\rm cut} < 1 \, .
\label{eq:profile_func}
\end{equation}
This profile function has some advantages such as
exact integrability and smoothness near the finite end point at $a_{\rm cut}$.
In this work, $m=2$ and $n=3$ are used.

In DJBUU, the dense medium effects are described by the mean fields obtained from
the relativistic Lagrangian density
consisting of nucleons, isoscalar (Lorentz scalar $\sigma$, Lorentz vector $\omega$), 
and isovector (Lorentz vector $\rho$) mesons;
\begin{eqnarray}
{\cal L} =&& \, \bar{\psi}[i \gamma_\mu \partial^\mu - (m_N + g_\sigma\sigma) - g_\omega \gamma_\mu\omega^\mu  \nonumber\\
&& - g_\rho \gamma^\mu \vec{\tau} \cdot \vec{\rho}^{\,\mu} -{e \over 2}\gamma_\mu (1+\tau^3)A^\mu]\psi \nonumber\\
&& + {1 \over 2}\left(\partial_\mu\sigma\partial^\mu\sigma - m_\sigma^2\sigma^2 \right) - {1 \over 3}a\sigma^3 - {1 \over 4}b\sigma^4 \nonumber\\
&& + {1 \over 2}m_\omega^2\omega_\mu\omega^\mu  + {1 \over 2}m_\rho^2\vec{\rho}_{\,\mu} \cdot \vec{\rho}^{\,\mu} \nonumber\\
&& - {1 \over 4}\Omega_{\, \mu \nu}\Omega^{\mu \nu} - {1 \over 4}\vec{R}_{\, \mu \nu} \cdot \vec{R}^{\, \mu \nu} -{1 \over 4} F_{\, \mu\nu}F^{\, \mu\nu}\, ,
\end{eqnarray}
where the over-arrow on $\vec{\rho}$ indicate the isospin vector nature of $\rho$ mesons and
field-strength tensors for the vector mesons ($\omega$ and $\rho$) and the electromagnetic field ($A_\mu$) are defined as
\bea
\Omega_{\, \mu\nu} &&= \partial_\mu\omega_\nu - \partial_\nu\omega_\mu \, \label{eq:omega_tensor},\\
\vec{R}_{\, \mu\nu} &&= \partial_\mu \vec{\rho}_{\mu} - \partial_\nu\vec{\rho}_{\nu} \, \label{eq:rho_tensor},\\
F_{\, \mu\nu} &&= \partial_\mu A_\nu - \partial_\nu A_\mu \, \label{eq:A_tensor}.
\eea

In the relativistic mean field approximation, a test particle propagates 
according to the classical equations of motion
 \bea
{d{\bf x}_\alpha\over dt} &=& {{\bf p}_\alpha\over E_\alpha}, \nonumber\\
{d{\bf p}_\alpha\over dt} &=& -\nabla V_\alpha^0 - {m_\alpha^*\nabla m_\alpha^* \over E_\alpha},
\label{eq:dxdp}
\eea
Here, $\alpha$ is the particle label,
$E_\alpha = \sqrt{{\bf p}_\alpha^2 + {m_\alpha^*}^2}$ is the energy, $V_\alpha^0$ is the vector
potential composed of $\omega$ and  $\rho^0$ vector meson mean fields,
and $m_\alpha^*$ is the effective mass in dense medium. 
For the nucleons, the effective mass is given by
$m_N^* = m_N - g_\sigma \sigma$ where $\sigma$ is the sigma meson mean field and 
$g_\sigma$ is the coupling constant. 
More detailed code description 
can be found in Ref.~\cite{MK16}.
For the comparison with the TCCP results, 
we are taking a particular parameter set (Set I) from Ref.~\cite{Liu:2001iz} 
as suggested by the transport code comparison project.
The mean field parameters and vacuum masses of nucleons and mesons are summarized 
in Table~\ref{table:MF_set}.
Following the TCCP procedures detailed in Ref.~\cite{Xu:2016lue},
we neglect the derivatives when solving the mean field equations
and only the time component of the vector meson fields are used.

At each time step, particles are sampled and paired
with other test particles which are geometrically closer than 
$d \le \sqrt{\hat{\sigma}/\pi}$.
In DJBUU, 
particles which have undergone scatterings are not allowed to decay
in the same time step, and they are not allowed to scatter further until they are 
sufficiently separated from their scattering partners.
Uncertainties caused by these constraints can be reduced by taking smaller time steps.

\section{Comparison with Transport Code Comparison Project}
\label{sec:DJBUU_comparison}

Many transport codes in BUU and QMD types have been developed for heavy ion collisions.
Main purpose of Transport Code Comparison Project (TCCP)
 is to have better predictions on the important physical quantities 
of HICs by reducing simulation uncertainties among different codes.
Main goal of this section is to validate DJBUU by comparing its results
with the TCCP results.

The project has already published
results for Au + Au collisions, box calculation for collisions
and box calculation for pion production~\cite{Xu:2016lue, Zhang:2017esm, Ono:2019ndq}.
Ideally, all codes should give the same results starting from
the same initial configuration. 
However, the TCCP
found that the numerical uncertainties among different codes reach up to 30$\%$. 
Because of the large uncertainties, the TCCP
 published other papers focused on 
collisions and Pauli blocking and pion production~\cite{Zhang:2017esm, Ono:2019ndq}.
They are also preparing a paper for the mean field dynamics in the box
calculation~\cite{MC19}.
Even though there are differences among the codes, 
the results from the project can be used to test the validity of 
the newly developed DJBUU code.
All the results below are obtained following the TCCP procedures and options
which are briefly described below.

For the heavy ion collisions ($^{197}$Au+$^{197}$Au),
we consider two different beam energies, $E_{\rm beam}=100A$ MeV (the B-mode in \cite{Xu:2016lue})
and $400A$ MeV (the D-mode in \cite{Xu:2016lue}). 
We use the the same initial conditions as in the TCCP
including the impact parameter fixed at $b=7$ fm.
We also consider the same three modes studied in TCCP
(i) only the mean fields are turned on without collisions (Vlasov), 
(ii) only collisions are turned on without the mean fields (Cascade),
and (iii) both the mean fields and the collisions are turned on (Full).
Only elastic collisions of nucleons are considered.
The included mean fields are $\sigma$, $\omega$ and $\rho^0$.
For the comparison with the TCCP, 
we focus on initialization, propagation, collision and final distribution.

For the infinite matter calculation (box calculation), we set the box size to be 20 fm and randomly distribute nucleons
to make the average density to be the nuclear saturation density (680 protons and 680 neutrons
in a cube with 20 fm edges).
In the momentum space, 
particle momenta are randomly distributed in the corresponding Fermi sphere
for two temperatures; $T=0$ MeV
and $T=5$ MeV. 
Only the collision and Pauli blocking effects
without the mean fields are considered in the box calculation.
Again, only elastic collisions of nucleons are considered and the protons and the neutrons have
the equal vacuum mass.
All results shown below
are calculated with 100 test particles and averaging over ten independent runs.

\subsection{Heavy-Ion Collisions}

\begin{figure}
\centering
\includegraphics[width=85mm]{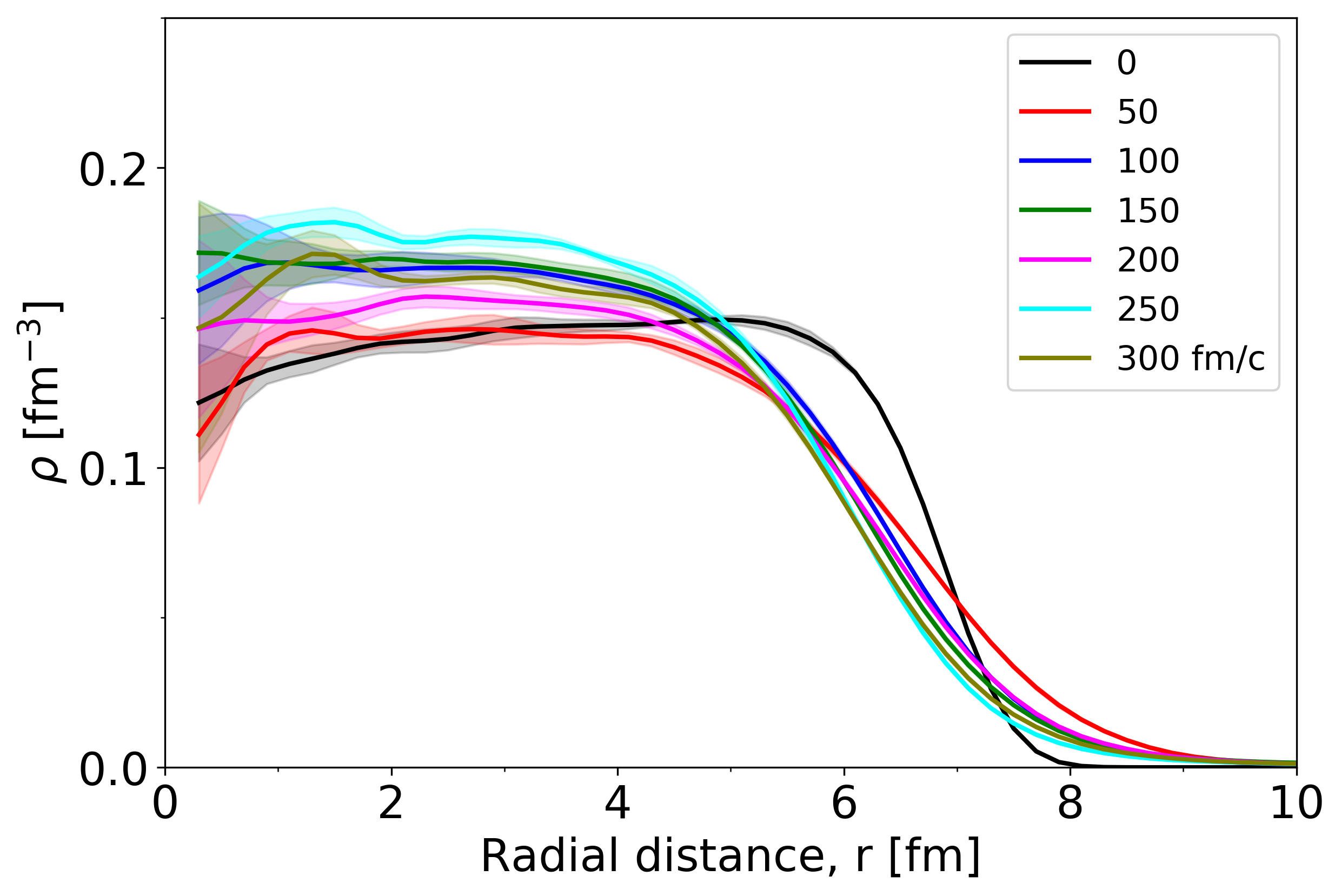}
\caption{Time evolution of the Au density profile in the intervals of 50 fm/$c$
at incident energy of 100$A$ MeV.
This is the average density of both projectile and target of Au.
Statistical uncertainty is shown as a band around the mean.}
\label{fig:stability}
\end{figure}

In this subsection, we compare our results 
with  those of the TCCP
on the time evolution of density distributions, collision rates, 
Pauli blocking factors, and momentum distributions.

One of the most important features that has to be checked in 
the transport simulation is the stability of nuclei.
Once a nucleus is generated, it should not collapse nor disperse away
unless it experiences a collision with other nucleus.
Fig.~\ref{fig:stability} shows the time evolution of
the averaged density profile of a stationary gold nucleus.
In the TCCP, Wood-Saxon form is used for the initial configuration of nuclei. 
However, in our simulation, we use relativistic Thomas-Fermi form
as it is more consistent with the mean-field dynamics.
The simulation has been performed with an extremely large impact parameter $b=20$ fm
so that the two nuclei won't collide.
Our results in Fig.~\ref{fig:stability} show that the density distributions of nuclei are oscillating.
However, even though the initial configuration is different, 
we confirm that the stability of stationary nuclei in DJBUU code is within the uncertainty of the
transport model comparison project. 

\begin{figure}
\centering
\includegraphics[width=85mm]{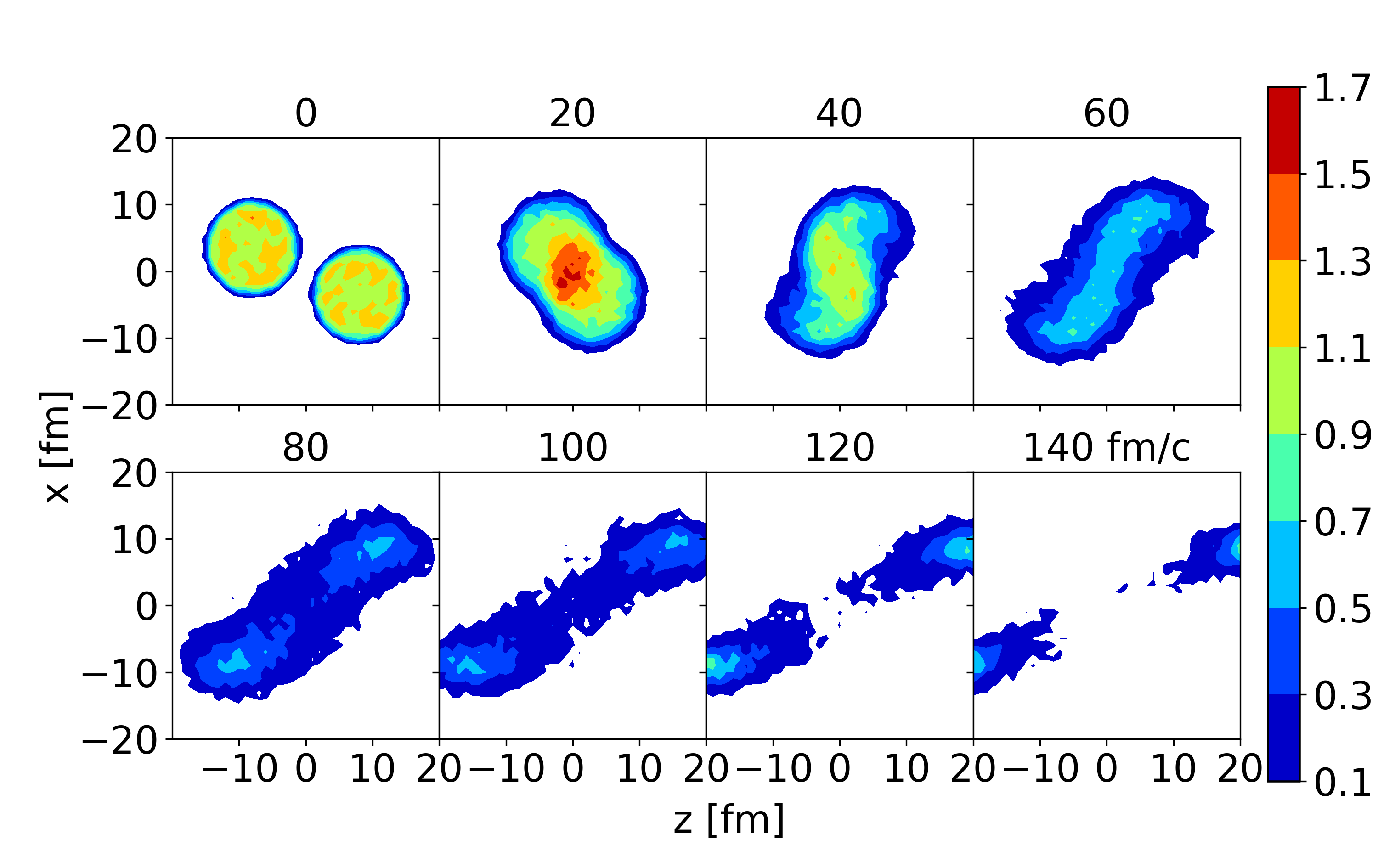}
\caption{Time evolution of density contours in Au + Au collision with an impact parameter $b=7$ fm
and incident beam energy 100$A$ MeV. 
Numbers on the top of each plot represent time in unit of fm/c. }
\label{fig:contour}
\end{figure}

For the processes with collisions, we take the impact parameter $b=7$ fm for Au+Au collisions.
Fig.~\ref{fig:contour} shows the evolution of nuclear density in a typical collision.
In the figure, we show the density contours 
in the $x-z$ plane at the 20 fm/$c$ time intervals in Au+Au collisions 
with incident energy at $100A$ MeV.
Here, $x$ is the direction of the impact parameter and $z$ is the beam direction.
In this particular example,
Coulomb interaction is not included and only elastic $NN$ scatterings are included.
Maximum density above $1.5\rho_0$ is reached near $t=20$ fm/$c$,
and the sideward flows are developed during $t=40 \sim 80$ fm/$c$ which is consistent with
the results in the TCCP study.

\begin{figure}
\centering
\includegraphics[width=85mm]{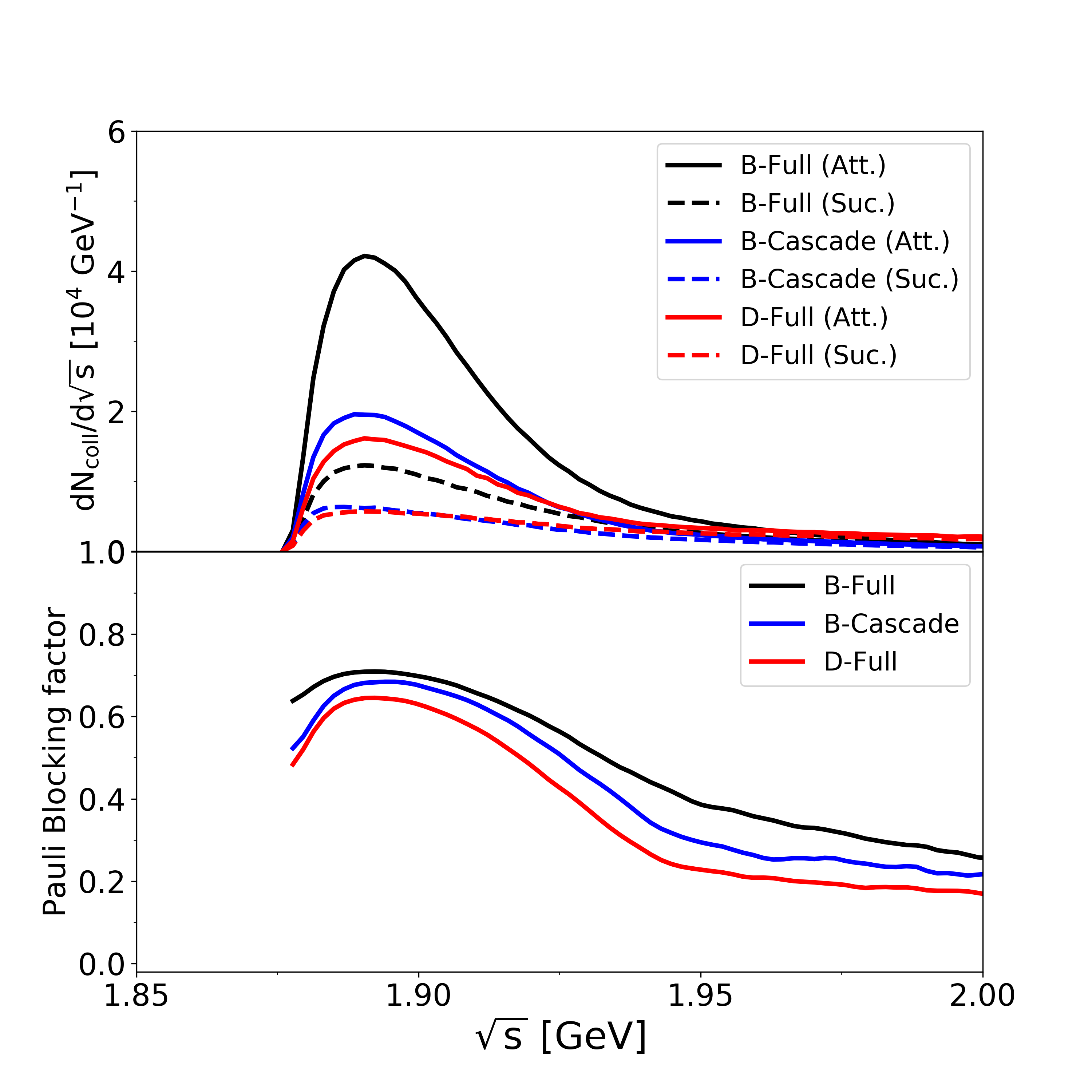}
\caption{Number of attempted and successful collisions (upper panel)
and Pauli blocking factor in Au+Au collisions (lower panel)
for two different incident energy 100$A$ MeV (B) and 400$A$ MeV (D).}
\label{fig:collision_pb}
\end{figure}

Following the TCCP procedure, we now check the successful collision rates and 
the Pauli blocking effects as a function of total energy in the center-of-mass 
frame for each collision. 
Even though these quantities are not directly detectable in experiments,
they are worth a close look to check the validity of the code.
In Fig.~\ref{fig:collision_pb}, number of total and successful collisions are shown in the 
upper panel, and the Pauli blocking factors, 
defined as the fraction of the aborted collisions
are shown in the lower panel. 
All quantities in the figure are integrated over the whole evolution time, and
only cascade and full mode simulations are plotted since collisions do not
occur in the Vlasov mode.
Even though the effective mass has to be used for the total energy 
in the center-of-mass frame, $\sqrt{s}=2\sqrt{{m_N^*}^2+p^2}$, 
vacuum mass is used for $\sqrt{s}$ in this plot to compare with other results of 
the transport code comparison project~\cite{Xu:2016lue}. 

In the figure, it is clearly seen that the collision number distribution has a peak
at $\sqrt{s} = 1.89$ GeV 
for $E_{\rm beam} = 100A$ MeV (B-mode)
which is slightly above the two nucleon threshold energy.
The peak is slightly shifted to a higher value for $E_{\rm beam} = 400 A$ MeV (D-mode)
because there are more nucleons with higher momentum.
The full mode with the mean fields at low energy (B-Full)  
has more collisions than those the B-Cascade mode without the mean fields,
or the D-Full mode with higher incident energy.
This indicates that the mean field facilitates collisions and the slightly lower number of
collisions for the D-mode reflects the fact that the total cross-section is a decreasing
function of $\sqrt{s}$ in this energy region.
The blocking factor is largest near the peak of the number of 
collisions because the phase space volumes of the occupied nuclei are largest  at the peak energy. 
The TCCP results for the collision numbers and the Pauli-blocking factor varies quite
significantly (see Figs.7 and 8 in Ref.~\cite{Xu:2016lue}. 
Our results are all well within the variation.

\begin{figure}
\centering
\subfigure[\, Transverse flow]{\label{fig:transflow}\includegraphics[height=68mm]{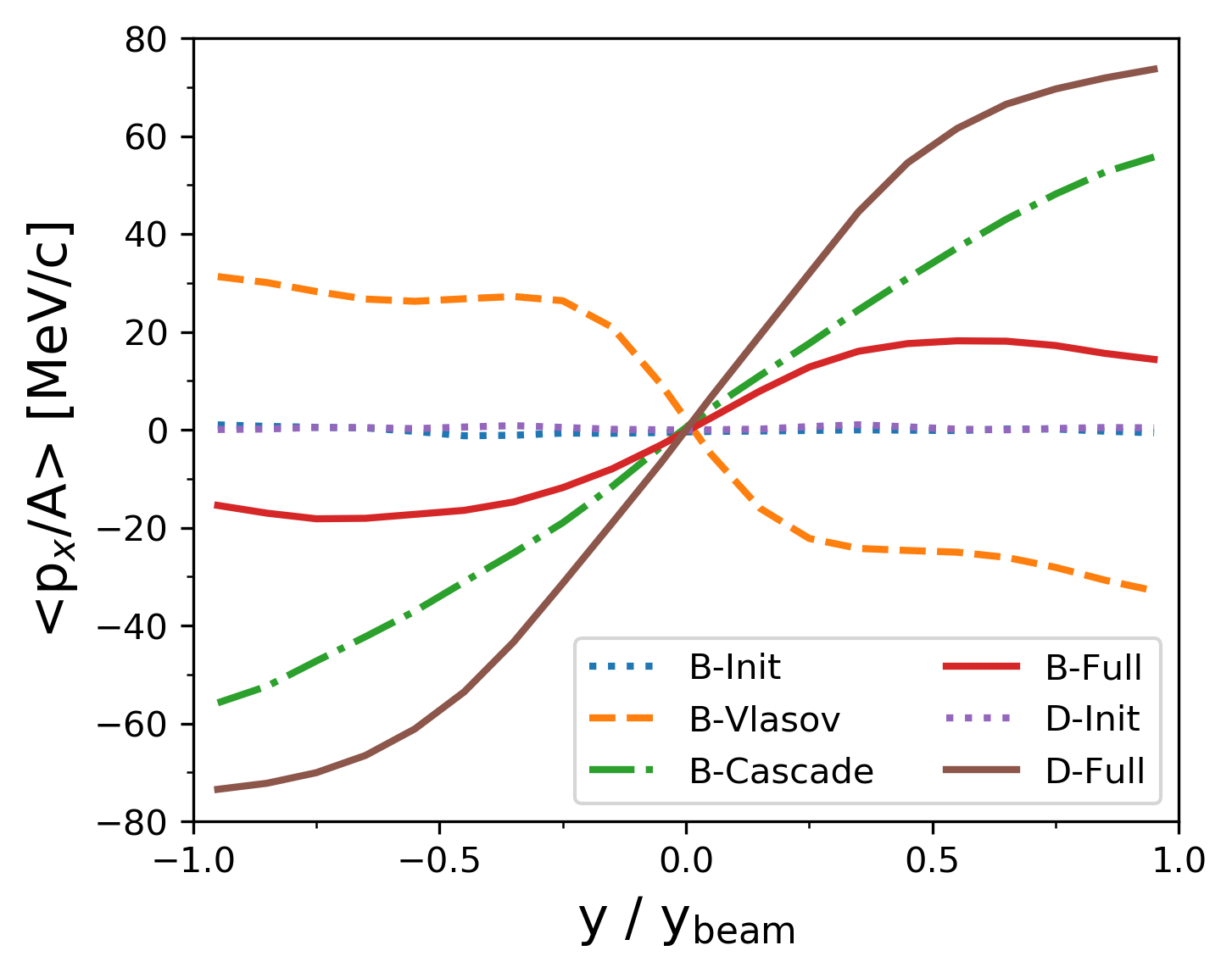}} \\
\subfigure[\, Rapidity distribution in B mode]{\label{fig:rapidity}\includegraphics[height=68mm]{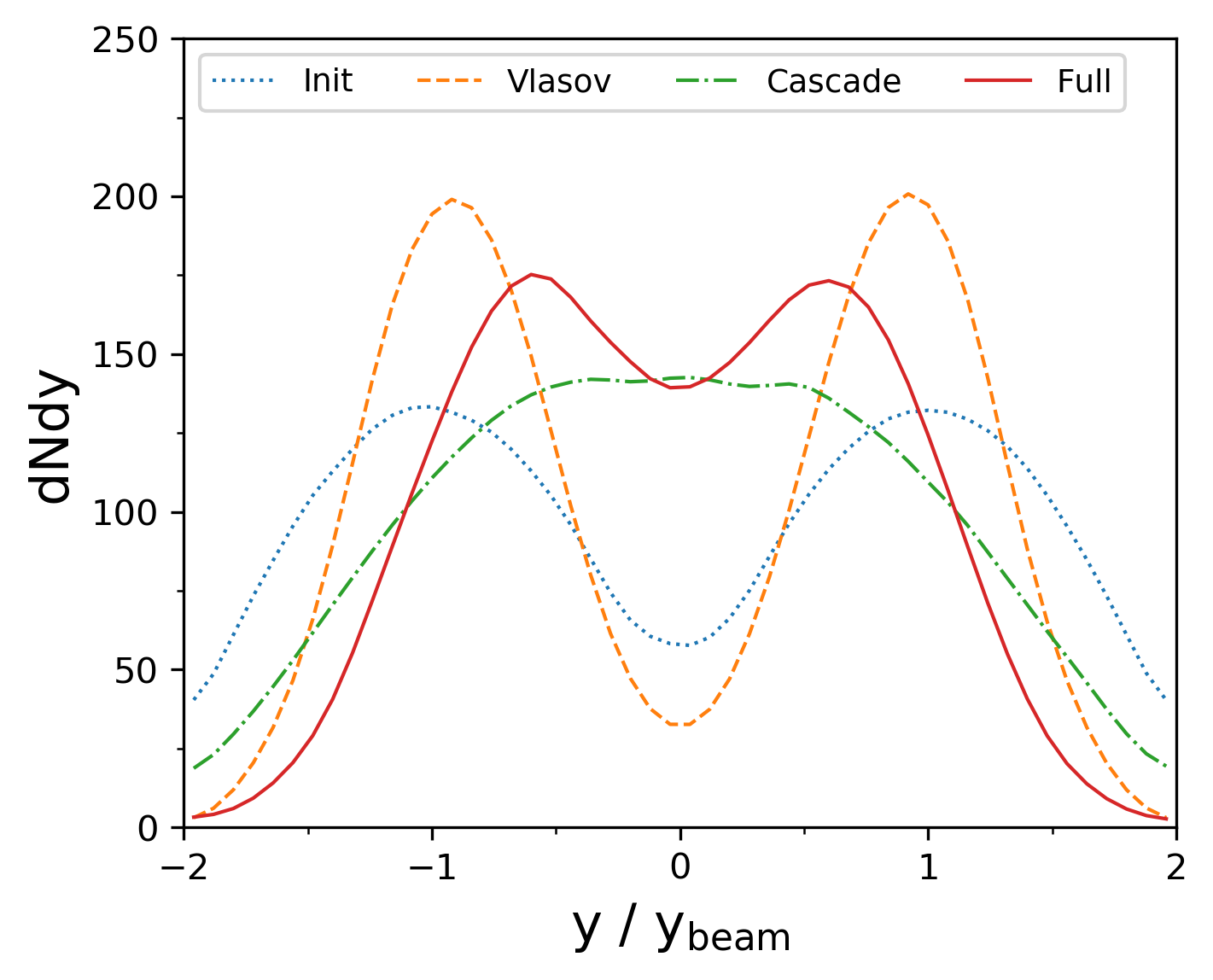}}
\caption{The particle distribution in impact parameter ($x$) and beam direction with respect to the reduced rapidity.
(a) Initial and final transverse flow for three different modes (Vlasov, Cascade and Full)
with two different beam energies (B: 100$A$ MeV and D: 400$A$ MeV).
(b) Initial and final rapidity distributions with $E_{\rm beam} = 100 A$ MeV (B). 
Here, the impact parameter $b=7$ fm.}
\label{fig:transflow}
\end{figure}

Having checked the overall collision dynamics, we now move on to observable results.
In heavy ion collisions, the final state momentum distribution encodes much information on the bulk
evolution. In the transverse plane, the anisotropic collective flow in the impact parameter direction
reflects how the original energy flow in the beam direction translates into the transverse pressure
due to interactions. In the longitudinal (beam) direction, the shape of the
rapidity distribution reflects how the longitudinal momentum transforms into transverse pressure.

To compare with results from other codes,
we generated events using the same initial conditions as in Ref.~\cite{Xu:2016lue}.
The average momentum in the $x$ direction at different rapidities 
are shown in Fig.~\ref{fig:transflow}(a). 
This particular observable is sensitive to the interaction between the spectator nucleons
and the participant nucleons.
As it should be, the
initial momentum distribution is almost uniform in the $x$ directions for both
the 100\,MeV beam energy (B-init) and the 400\,MeV beam energy (D-init).
However, final momentum distributions are strongly influenced by the presence of 
the mean fields and scatterings.
If the scatterings are turned off, then higher baryon density generates higher $\sigma$ mean field
which provides more attraction towards the spectator nucleons. 
On the other hand, if the mean fields are turned off, then higher 
baryon density implies higher rates of scatterings between the spectators and the
participants which provides effective pressure away from the spectators.
This effect is most clearly seen in the low energy collisions at $E_{\rm beam}=100A$\,MeV
because the spectators are slower to move away from the collision region.
One can see that the Vlasov mode (B-Vlasov) and the Cascade mode (B-Cascade) clearly exhibit
opposite sign slopes.
In the full mode (B-Full), the effect of scattering is larger than 
that of the mean fields causing a positive, but more gentle, slope at the mid-rapidity region.
At $E_{\rm beam}= 400 A$ MeV, the scattering effect is even stronger.
We note that the attraction caused by scalar mean fields 
and the repulsion caused by vector mean fields balance 
at $E_{\rm crit}$ ($\approx 140 A$ MeV in Ref.~\cite{Xuemin90}), 
and the mean field effect is attractive for $E_{\rm beam} < E_{\rm crit}$, 
but repulsive for $E_{\rm beam} > E_{\rm crit}$. 

\begin{figure}
\centering
\includegraphics[width=80mm]{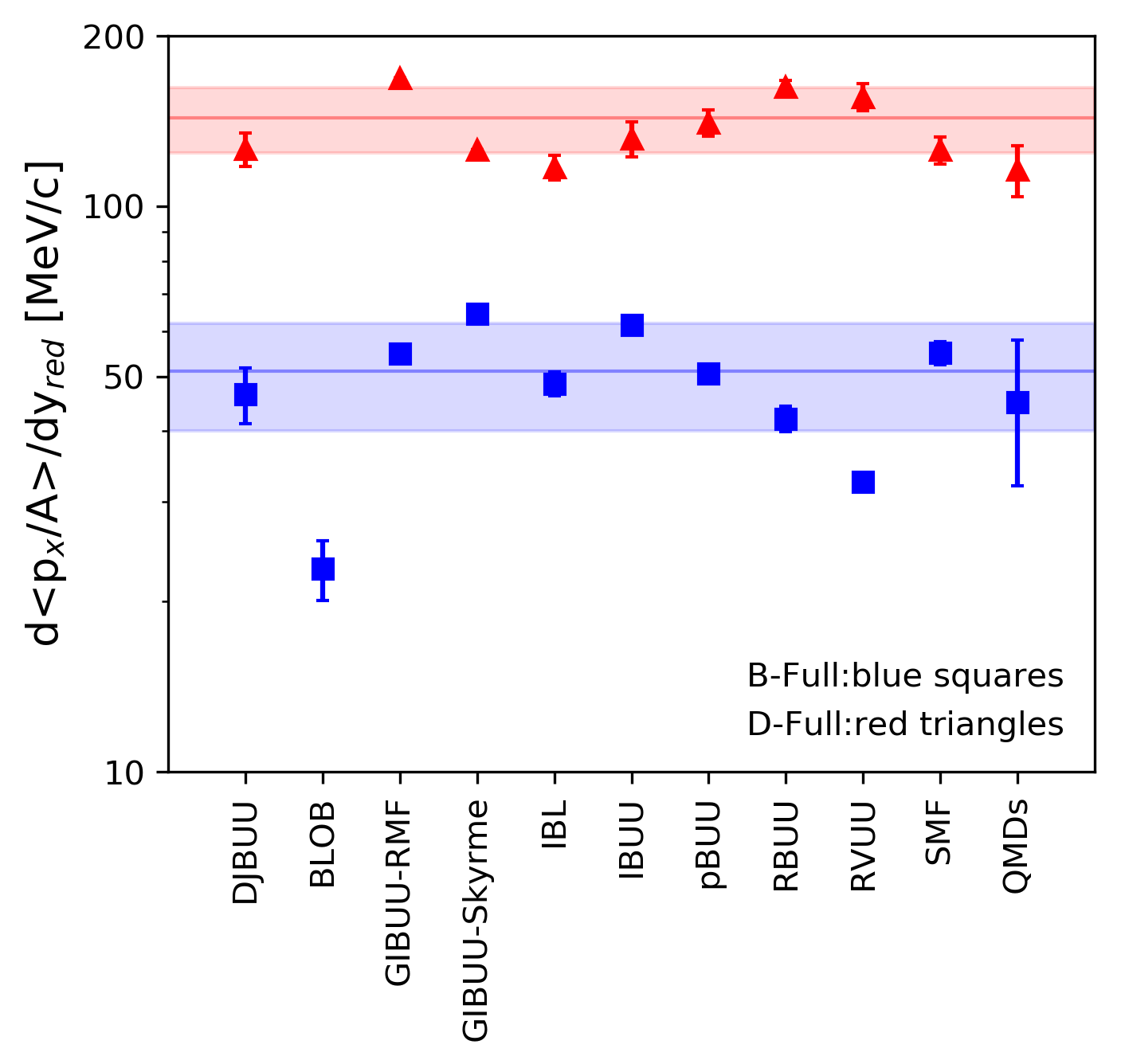}
\caption{Slope parameters of DJBUU,  nine BUUs and QMDs at mid-rapidity.
Two shaded regions are mean and standard deviation of nine BUUs at
beam energy 100 (blue) and 400$A$ MeV (red).}
\label{fig:slope_param}
\end{figure}

\begin{table*}
\begin{center}
\begin{threeparttable}
\begin{tabular}{c|ccc|ccc}
\hline
\hline
&&{Pauli blocking}& &&  {Slope parameter}& [MeV/$c$]  \\
&DJBUU & BUUs & QMDs  &DJBUU & BUUs & QMDs \\
\hline
B-Cascade & 0.677	&0.65 $\pm$ 0.129	& 0.51 $\pm$ 0.212	& \\
B-Full & 0.700	& 0.75 $\pm$ 0.124	& 0.70 $\pm$ 0.136	& $46.5\pm5.3$ & 51$\pm$ 11 &45 $\pm$ 13\\
D-Full & 0.630	& 0.63 $\pm$ 0.145	& 0.55 $\pm$ 0.138	& $126.1 \pm 8.7$ & 143 $\pm$ 19 & 116 $\pm$ 12\\
\hline
\hline
\end{tabular}
\caption{Pauli blocking factor at $\sqrt{s} = 1.9$ GeV,
and the mean transverse flow of DJBUU, BUUs and QMDs
at B-Cascade or Full (100$A$ MeV with only collisions or both collisions and the mean field)
 and D-Full (400$A$ MeV).
The mean flow of BUUs and QMDs are from Ref.~\cite{Xu:2016lue}.}
\label{table:djbuu_tccp_HIC}
\end{threeparttable}
\end{center}
\end{table*}

In Fig.~\ref{fig:rapidity}, 
the rapidity distributions with $E_{\rm beam}=100 A$ MeV (B-mode) are summarized.
Initially projectile and target sit at the reduced rapidity $y/y_{\rm beam}=\pm1$. 
Positive (negative) rapidity corresponds to the projectile (target).
The peaks in the final distribution of Vlasov mode are shifted toward 
the center because of the attractive effect in B-Vlasov model. 
Without the mean fields (B-Cascade mode), the distribution fills the mid-rapidity region 
because the stopping.
In the B-Full mode where both mean field and collision effects are considered,
the final distribution is between those of B-Vlasov and B-Cascade.
These results are all consistent with those presented in the TCCP study.

Fig.~\ref{fig:slope_param} compares the slope parameter
which is a linear fit of transverse flow in the rapidity range
$| y/y_{\rm beam}| < 0.38$. 
In the TCCP study,
the mean and standard deviation of slope parameter
was $51\pm11$ MeV/$c$ at $100A$ MeV
and $143\pm19$ MeV/$c$ at $400A$ MeV
among 9 participating BUU codes.
The QMDs had  $45\pm13$ MeV/$c$ at $100A$ MeV
and $116\pm12$ MeV/$c$ at $400A$ MeV.

For the D-Full mode, the DJBUU result ($126.1 \pm 8.7$ MeV/c) is 
somewhat lower than that of other relativistic BUU codes (GIBUU-RMF, RBUU, and RVUU).
For the B-Full mode, the DJBUU result ($46.5 \pm 5.3$ MeV/c) is
consistent with others.
This could be due to the differences in 
the mean field calculations among the relativistic codes.
Unfortunately, those differences were not extensively explored in previous studies.
Nevertheless, our results are all within the uncertainties of the overall TCCP values.

The comparison results are summarized in
the Table~\ref{table:djbuu_tccp_HIC}.
In summary, the DJBUU results are 
consistent with those in the TCCP within the model uncertainties.

\subsection{Infinite Dense Matter}
Because of the differences in the implementation of the transport simulations for HICs,
the transport code comparison project suggested box calculations for checking three important ingredients
in the transport code: collisions and Pauli blockings, mean field dynamics and pion production
Ref.~\cite{Zhang:2017esm}.
In this work, for the low temperature simulations, 
we focus on the collisions and blockings because pion production is negligible at low temperature.
In this section, we compare our results with those in the second TCCP paper 
Ref.~\cite{Zhang:2017esm}.

\begin{figure}
\centering
\subfigure[\, CT0]{\label{fig:CT0}\includegraphics[height=55mm]{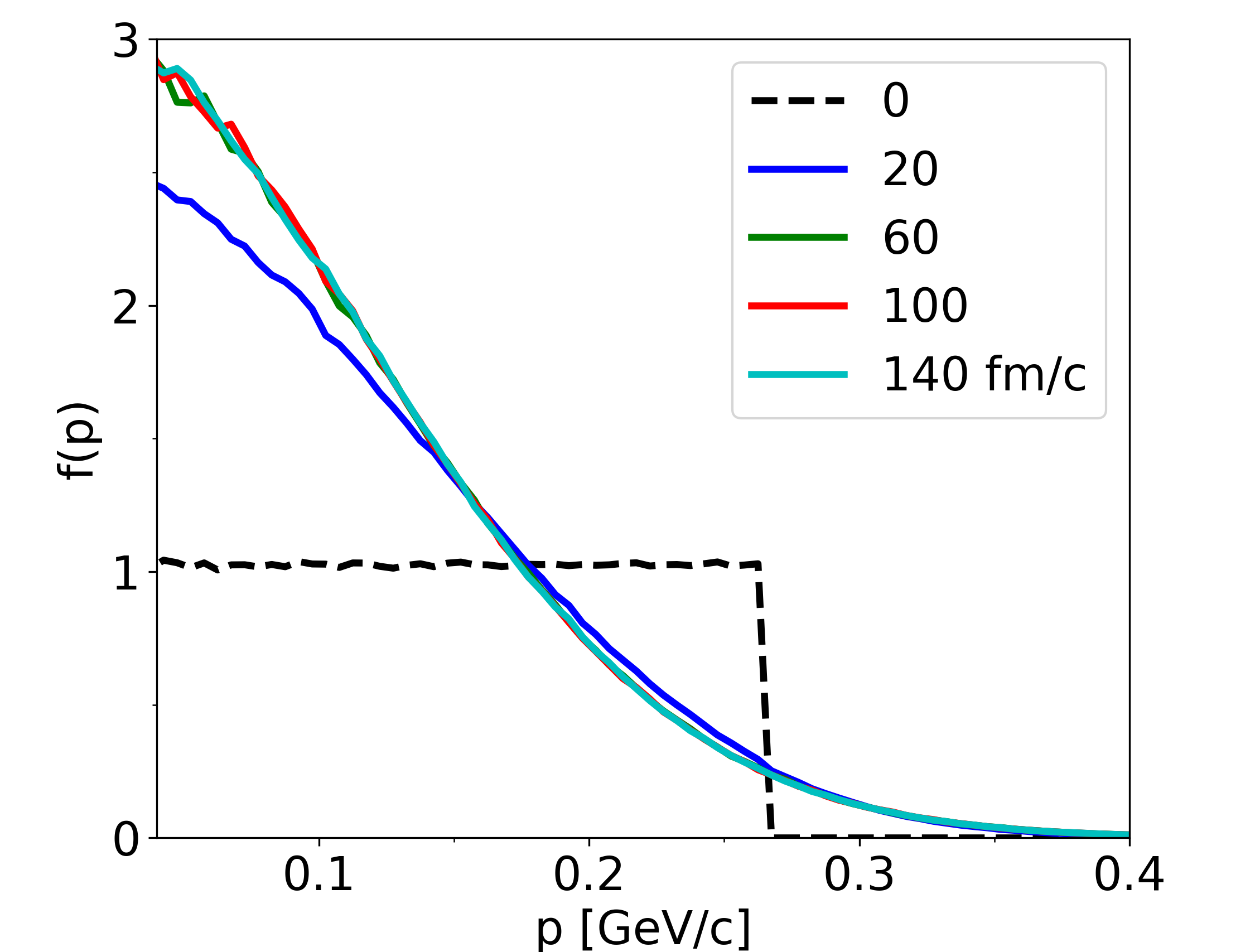}}\\
\subfigure[\, CT5]{\label{fig:CT5}\includegraphics[height=55mm]{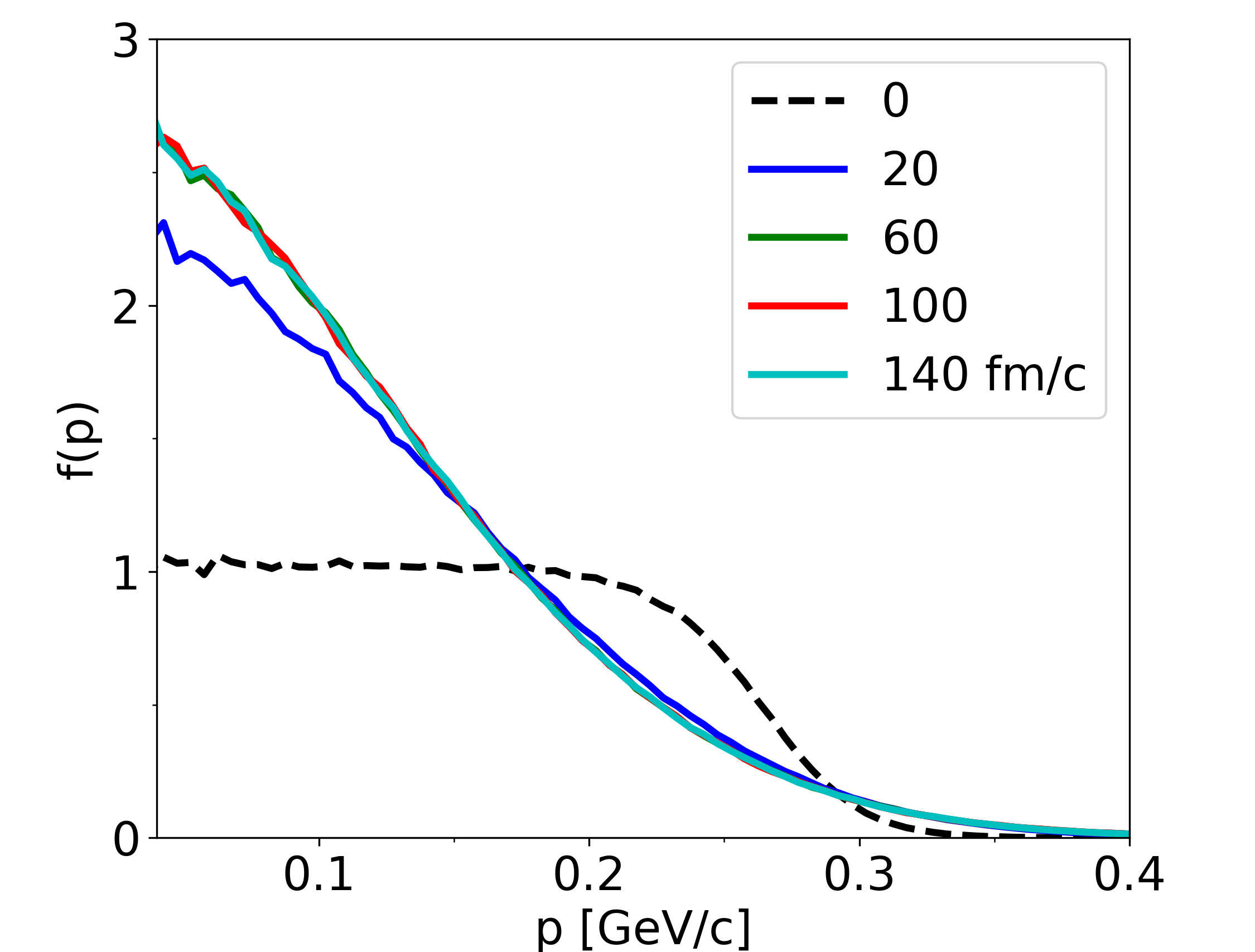}}
\caption{Momentum distributions at time $t=0, 20, 60, 100$ and 140 fm/$c$ for the cascade mode C (without Pauli blocking)
with $T=0$ MeV (a) and 5 MeV (b).}
\label{fig:CT_dist}
\end{figure}

For the box calculation, Ref.~\cite{Zhang:2017esm}
suggested two collision modes (C, CB) 
for two temperatures (T0, T5), 
and two Pauli blocking options (OP1, OP2) for CB.
Here, the mode C is a cascade mode without the mean fields and the Pauli
blocking, and
the mode CB is a cascade mode without the mean fields. 
T0 and T5 correspond to $T=0$ MeV and $T=5$ MeV, respectively. 
The option OP1 is with the collision and blocking methods intrinsic to DJBUU
as explained in Section~\ref{sec:DJBUU}.
The option OP2 is with the reference criteria for both collisions and blocking provided
by the TCCP for comparison in which the Pauli blocking is always calculated
with the initial thermal distribution regardless of the local environment of the particle
at the given time.
In total, six sets of calculations are carried out
as suggested by the TCCP: 
they are denoted as CT0, CT5, CBOP1T0, CBOP1T5, CBOP2T0, and CBOP2T5.

\begin{figure}
\centering
\includegraphics[width=95mm]{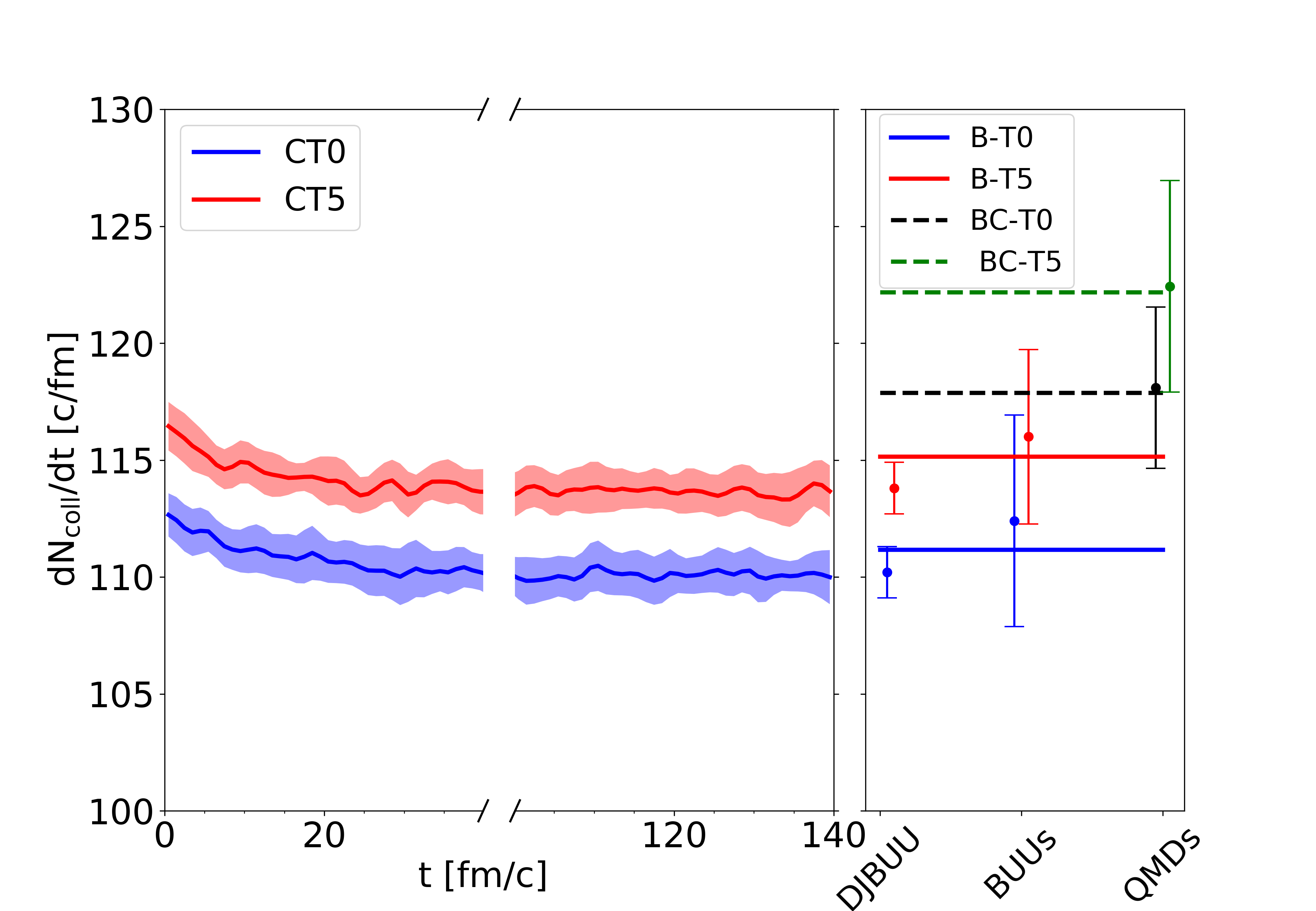}
\caption{
(Left) Time evolution of collision rate $dN_{\rm coll} / dt$ without Pauli blocking at
 $T=0$ (CT0) and 5 MeV (CT5).
 (Right) Averaged collision rate with time interval from 60 to 140 fm/$c$ for DJBUU, BUUs and QMDs.
 The straight solid and dashed lines represent the reference values from relativistic Boltzmann 
 at $T=0$ (B-T0) and 5 MeV (B-T5)
 and relativistic basic cascade code results
 at $T=0$ (BC-T0) and 5 MeV (BC-T5).}
\label{fig:dNdt}
\end{figure}

In Figs.~\ref{fig:CT0} and \ref{fig:CT5}  we show the momentum distributions 
at $t=0,20,60,100$, and 140 fm/$c$ with $T=0$ (T0) and  5 MeV (T5), respectively.
Even though initial momenta of particles are distributed according 
to the Fermi-Dirac distributions for both temperatures,
the final distributions of the momentum are expected to follow
the classical Boltzmann distributions due to the diffusion intrinsic to the coarse graining
procedure to calculate the phase densities~\cite{Xu:2016lue, Abe:1995yw}.
This numerical artifact was also observed in other models.
In our simulation, the fitted temperatures of the final distributions, 
with the assumption of the relativistic Boltzmann distribution,
are $T_B=14.355$ and 15.399 MeV for T0 and T5, respectively. 
These values are very close to the values obtained in the TCCP:
$T_B=14.284$ and 15.364 MeV for T0 and T5, respectively~\cite{Zhang:2017esm}.

Fig.~\ref{fig:dNdt} shows the time evolution of collision rate, 
$dN_{\rm coll}/dt$, for the mode C (without Pauli-blocking) with the $1\,\sigma$ uncertainties.
The initial collision rates are 112.8 and 116.8  for $T=0$ and 5 MeV, respectively.
One can compare these values with the reference values in the
TCCP~\cite{Zhang:2017esm}: 114.0 or 115.2 for $T=0$ and
117.8 or 119.0 for $T=5$ MeV. 
Note that they obtained two reference values for each temperature by changing the time step
; one is constant time step and the other is time dilation factor
Around $t=40$ fm/$c$ in Fig.~\ref{fig:CT_dist}, the momentum distributions become
Boltzmann likely distributions. 
Hence, after $t=40$ fm/$c$, we expect that the system reach equilibrium and the collision
rates saturate. 
In our simulation, the saturated collision rates averaged over time 
from 60 to 140 fm/$c$ are 110.2 and 113.8 for T0 and T5, respectively.

On the right panel of the Fig.~\ref{fig:dNdt} shows
collision rates of DJBUU and other transport codes (BUUs and QMDs).
The horizontal lines are the reference values for two temperatures $T=0$ and 5 MeV.
The reference values labeled with B
comes from evaluating the equilibrium collision rates using Boltzmann distributions.
The reference values labeled with BC
comes from calculating the collision rates in the `basic cascade' simulations in which only
the collision pairs at each time step are counted without actually colliding them.
Most of BUU types including DJBUU 
are close to the value of relativistic Boltzmann calculation
while most of QMDs are close to the relativistic basic cascade.

\begin{figure}
\centering
\includegraphics[width=90mm]{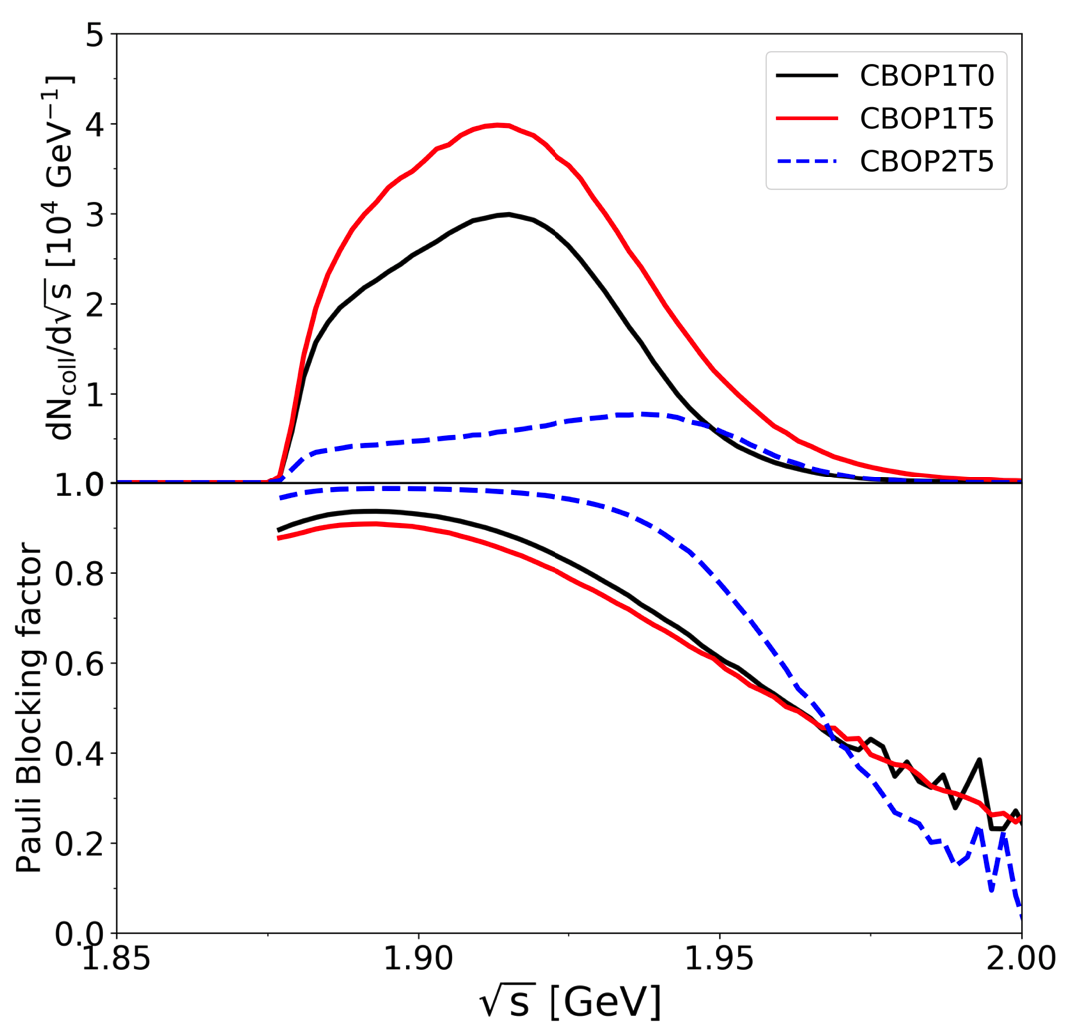}
\caption{(Left panel) Successful collision rate as a function of the center-of-mass energy for
$T=0$ (CBOP1T0) and 5 MeV (CBOP1T5).
(Right panel) Center-of-mass energy distribution of averaged Pauli blocking factors
defined as 1-(successful collisions / attempted collisions).}
\label{fig:box_pb}
\end{figure}

The successful collision rate and Pauli blocking factor in DJBUU (OP1) are
shown in Fig.~\ref{fig:box_pb}.
The successful collisions are peaked around 1.92 GeV
while the attempted collisions are peaked at slightly lower energy (not on the figure).
The time averaged Pauli blocking factor as a function of energy is plotted in the right panel. 
Again, the TCCP study found that these results vary quite substantially among the tested codes
just as they were in the Au-Au collision study.
The DJBUU results are certainly within the variation show in in Fig.~5
in Ref.~\cite{Zhang:2017esm}.
The dashed line in the figure corresponds to the OP2 in which the Pauli blocking
is always calculated with the initial Fermi-Dirac distribution with $T=5$ MeV.

In Table~\ref{table:coll_rate}, we summarize the successful collision rates in box calculations with Pauli blocking
for initial temperatures at $T=0$ and 5 MeV. 
We checked the collision rates for the first time step (1st $\Delta$t)
and the rate averaged over time interval 60-140 fm/$c$ as in the TCCP.
In the comparison project, most of QMD families have large collision rates, $20 \sim 40$ $c$/fm,
but BUU types have smaller rates, $10 \sim 20$ $c$/fm except for pBUU.
The collision rates of DJBUU (CBOP1) are consistent with the BUU types.
This assures that the collisions and blockings are working properly in DJBUU. 
With the ideal Pauli blocking option at $T=0$ (CBOP2T0), collision rates are zero 
since all collisions must be blocked.
This is because in OP2 at $T=0$ the Fermi-Dirac distribution is either 1 or 0.
For ideal option at $T=5$ MeV (CBOP2T5), collisions rates are slightly 
lower than the theoretical estimation by the TCCP,
 3.5 $c$/fm (relativistic cases) but acceptable.

\begin{table}
\begin{center}
\begin{threeparttable}
\begin{tabular}{c|cc|c|c}
\hline
\hline
& \multicolumn{2}{c}{DJBUU} & {BUUs} &{QMDs} \\
&OP1T0&OP1T5 &OP1T5&OP1T5\\
\hline
1st $\Delta$t& 11.217 & 16.372   & 4.2-23.12 &  3.34 - 38.83\\
t$_{\rm avg.}$ & 11.161 & 16.077   & 4.2-22.67 & 3.34 - 40.91\\
\hline
\hline
\end{tabular}
\caption{Successful collision rates $dN^{\rm suc}_{\rm coll}/dt$
with Pauli blocking for
four options in the unit of c/fm.
The row marked with 1st $\Delta$t has the rate for the first time step,
while the row marked with t$_{\rm avg.}$ has the rate
averaged over time interval 60-140 fm/$c$.
The minimum and maximum collision rates are
 taken from the Fig. 8. in Ref.~\cite{Zhang:2017esm}.}
\label{table:coll_rate}
\end{threeparttable}
\end{center}
\end{table}

In this section, we have compared our results for collisions 
and blocking with the TCCP results.
We have also tested other physical quantities, such as the pion production suggested by the project 
and found that our results are consistent other results. We can conclude that DJBUU has 
successfully passed the infinite matter test.

\section{The Extended Parity Doublet Model}
\label{sec:pdm}

Up to now, we have applied our model to the idealized cases
to test the inner workings of the
code. With the confidence gained by testing DJBUU against the TCCP tests,
we now would like to apply DJBUU
to realistic heavy-ion collisions and test a specific physics model.
The physics model we chose to test is the Extended Parity Doublet model 
(EPDM)~\cite{Motohiro:2015taa}.
The motivation for implementing this model in DJBUU is to see 
how the observable from HICs depends on the chiral invariant mass.
In this subsection, we briefly introduce the Extended Parity Doublet Model.

\begin{table*}
\begin{center}
\begin{tabular}{ l| l l l l | l l l l}
\hline
\hline
 & &  \multicolumn{2}{c}{$K = 215$ MeV} & & & \multicolumn{2}{c}{$K = 240$ MeV} & \\
 $m_0$       $~~~~~~~~$   & $600$  $~~~~~~~$&   $700$    $~~~~~~~$   & $800$   $~~~~~~~$  & $900$ $~~~~~~~$ & $600$ $~~~~~~~$ &   $700$     $~~~~~~~$  & $800$    $~~~~~~~$ & $900$ \\
\hline
 $g_1$              &  $14.836$   & $14.1708$ & $13.3493$ & $12.3293$ &  $14.836$ &  $14.1708$ & $13.3493$ & $12.3293$\\
 $g_2$               & $8.42735$    & $7.76222$ & $6.94073$ & $5.92073$  & $8.42735$ &  $7.76222$ & $6.94073$ & $5.92073$\\
 $g_{\omega}$        & $8.90217$ & $7.05508$ & $5.47079$ & $3.38862$  & $9.13193$ &  $7.30465$ & $5.65978$ & $3.52185$\\
 $g_{\rho}$            & $3.97462$  & $4.07986$ & $4.15669$ & $4.22091$ & $3.92698$ &  $4.06502$ & $4.14894$ & $4.21785$\\
 $\bar{\mu}^2/f_\pi^2$   & $23.3772$ & $20.9799$ & $13.3463$ & $2.50198$ & $21.8212$  &  $18.8421$ & $11.6928$ & $1.5374$\\
 $\lambda$            & $42.3692$   & $38.921$  & $26.1283$ & $6.673$ & $39.3674$  &  $34.5841$ & $22.5779$ & $4.38835$\\
 $\lambda_6 f_\pi^2$    & $16.7901$ & $15.7393$ & $10.5802$ & $1.96915$ & $15.3444$ &  $13.5401$ & $8.68327$ & $0.649073$\\
 $m_\sigma$             & $413.612$ & $384.428$ & $324.007$ & $257.583$  & $411.299$ &  $385.805$ & $330.44$  & $269.255$\\
\hline
\hline
\end{tabular}
\caption{Parameter sets used in this work with different compressibility: $K = 215$ and $K = 240$~MeV.
$m_0$ and $m_\sigma$ are in MeV.
The parameter sets which are fixed to fit nuclear matter properties for given compressibility $K$ and $m_0$
~\cite{Shin:2018axs}.}
\label{table:param_set1}
\end{center}
\end{table*}

The Lagrangian for EPDM constructed in Ref.~\cite{Motohiro:2015taa} is given by
\begin{eqnarray}
{\cal L} &&= \bar{\psi}_1 i \gamma_\mu \partial^\mu \psi_1 + \bar{\psi}_2 i \gamma_\mu \partial^\mu \psi_2 + m_0 \left( \bar{\psi}_2 \gamma_5 \psi_1 - \bar{\psi}_1 \gamma_5 \psi_2 \right) \nonumber\\
&& + g_1\bar{\psi}_1 \left( \sigma + i \gamma_5 \vec{\tau} \cdot \vec{\pi}\right) \psi_1 + g_2\bar{\psi}_2 \left( \sigma - i \gamma_5 \vec{\tau} \cdot \vec{\pi}\right) \psi_2 \nonumber\\
&& -g_{\omega NN} \bar{\psi}_1 \gamma_\mu \omega^\mu \psi_1 -g_{\omega NN} \bar{\psi}_2 \gamma_\mu \omega^\mu \psi_2 \nonumber\\
&& -g_{\rho NN } \bar{\psi}_1 \gamma_\mu \vec{\rho}^{\,\mu} \cdot \vec{\tau} \psi_1 -g_{\rho NN} \bar{\psi}_2 \gamma_\mu \vec{\rho}^{\,\mu} \cdot \vec{\tau} \psi_2 \nonumber\\
&& -e \bar{\psi}_1 \gamma^\mu A_\mu {{1-\tau_3}\over 2} \psi_1 -e \bar{\psi}_2 \gamma^\mu A_\mu {{1-\tau_3} \over2} \psi_2 + {\cal L}_M\, ,
\end{eqnarray}
where the right-handed and the left-handed components of the
baryon fields $\psi_1$ and $\psi_2$ transform as
\begin{eqnarray}
&&\psi_{1R} \rightarrow R\psi_{1R}, \quad \psi_{1L} \rightarrow L\psi_{1L}\, , \nonumber \\
&&\psi_{2R} \rightarrow L\psi_{2R}, \quad \psi_{2L} \rightarrow R\psi_{2L}\, ,
\end{eqnarray}
where $R$ is an element of the $SU(2)_R$ chiral symmetry group and
$L$ is an element of the $SU(2)_L$ chiral symmetry group.
Here $m_0$ represents the chiral invariant mass.

The mesonic part of the Lagrangian reads
\begin{eqnarray}
{\cal L}_M &&= {1 \over 2} \partial_\mu \sigma \partial^\mu \sigma + {1\over2} \partial_\mu \vec{\pi} \cdot \partial^\mu \vec{\pi} \nonumber\\
&& -{1 \over 4} \Omega_{\mu\nu} \Omega^{\mu\nu} -{1 \over 4} \vec{R}_{\mu\nu} \cdot \vec{R}^{\mu\nu} -{1\over4} F_{\mu\nu} F^{\mu\nu} \nonumber \\
&& + {1 \over2} \bar{\mu}^2\left( \sigma^2 + \vec{\pi}^2 \right) - \frac{\lambda}{4} \left( \sigma^2 + \vec{\pi}^2 \right)^2 + {1\over6}\lambda_6 \left( \sigma^2 + \vec{\pi}^2 \right)^3  \nonumber\\
&&+ \epsilon \sigma + {1 \over2}m_\omega^2 \omega_\mu \omega^\mu  + \frac{1}{2}m_\rho^2 \vec{\rho}_\mu \cdot \vec{\rho}^{\,\mu}\, ,
\end{eqnarray}
where $\Omega_{\mu\nu}, \vec{R}_{\mu\nu}$ and $F_{\mu\nu}$ are in Eqs.~(\ref{eq:omega_tensor})-(\ref{eq:A_tensor}).

The collective meson field
$M = \sigma + i\boldsymbol{\tau}\cdot\boldsymbol{\pi}$
transforms as
\begin{equation}
M \to L M R^\dagger \,.
\end{equation}
We note here that the pion mass $m_\pi$, $\sigma$ meson mass $m_\sigma$, and pion decay constant $f_\pi$
can be related to the parameters $\lambda$, $\bar{\mu}^2$ and $\lambda_6$ in vacuum:
\begin{eqnarray}
m_\pi^2&=&\lambda\sigma_0^2-\bar\mu^2-\lambda_6\sigma_0^4\, ,\nonumber \\
m_\sigma^2&=&3\lambda\sigma_0^2-\bar\mu^2-5\lambda_6\sigma_0^4\, ,\nonumber \\
f_\pi &=& \sigma_0\, ,
\end{eqnarray}
with $m_\pi = 138$ MeV, $f_\pi = 93$ MeV and $\sigma_0 = f_\pi$ the vacuum expectation value of the $\sigma$ field.
The mass of the $\sigma$ meson in this work is treated as a free parameter, while 
the masses of $\omega$ and $\rho$ meson are set to 
$m_\omega = 783$ MeV and $m_\rho = 776$ MeV.

We now make the mean field approximation 
by replacing the $\sigma, \omega$ and the $\rho$ field by their mean fields 
$\sigma\rightarrow \bar{\sigma}$,
$\omega_\mu\rightarrow \delta_{\mu 0}\bar{\omega}_0$, 
and $\rho_{i \mu}\rightarrow \delta_{i3}\delta_{\mu 0}\bar{\rho}_0^3$.
The equations of motion (EoM) for the stationary mean fields 
$\tilde{\sigma} = \bar{\sigma} - \sigma_0$,
$\bar{\omega}$, $\bar{\rho}$ and $\bar{A}_0$ read
\begin{eqnarray}
\left(-\vec{\nabla}^2 + m_\sigma^2 \right) \tilde{\sigma}(\vec{x}) &=&
- \bar{N}(\vec{x})N(\vec{x}) \left.\frac{\partial\, m_N( \tilde{\sigma})}{\partial \tilde{\sigma} }
\right|_{\tilde{\sigma}= \tilde{\sigma}(\vec{x})}\nonumber \\
&& +(-3f_\pi \lambda + 10f_\pi^3\lambda_6) \tilde{\sigma}(\vec{x})^2 \nonumber \\
&&+(-\lambda + 10f_\pi^2\lambda_6) \tilde{\sigma}(\vec{x})^3\nonumber \\
&& +5f_\pi\lambda_6 \tilde{\sigma}(\vec{x})^4
+ \lambda_6 \tilde{\sigma}(\vec{x})^5\,, \label{eq:EoMsig} \\
\left(-\vec{\nabla}^2 + m_\omega^2 \right) \bar{\omega}(\vec{x}) &=& g_{\omega NN}N^\dagger(\vec{x}) N(\vec{x})\,, \label{eq:EoMome}\\
\left(-\vec{\nabla}^2 + m_\rho^2 \right)\bar{\rho}(\vec{x}) 
&=& g_{\rho NN}N^\dagger(\vec{x})\tau_3  N(\vec{x})\,, \label{eq:EoMrho}\\
-\vec{\nabla}^2 \bar{A}_0(\vec{x}) \rangle &=& e N^\dagger(\vec{x})  \frac{1-\tau_3}{2} N(\vec{x})\,. \label{eq:EoMphoton}
\end{eqnarray}
Currently, only the time component of $\omega, \rho$ and $A$ are included
and the effect of the Laplacian term is included only for the 
electromagnetic potential $A_0$.
The mass eigenstates
are obtained by diagonalizing the mass matrix
\be
m_\pm = {1 \over 2} \left({\sqrt{(g_1 + g_2)^2 \bar{\sigma}^2 + 4m_0^2} 
\mp (g_1 -g_2)\bar{\sigma}}\right)\, .
\label{eq:mass}
\ee
The nucleon mass is $m_N = m_+$ since they have positive parity.
Its negative-parity partner has $m_-$.
Note that $\bar{\sigma} = \tilde{\sigma} + \sigma_0$
is the in-medium average that depends on the environment.

Using the nucleon mass, meson masses and pion decay constant,
one can determine meson coupling constants, $g_1, g_2, g_\omega, g_\rho$ and 
parameter $\lambda, \bar{\mu}^2$ and $\lambda_6$.
The nuclear matter properties used to fix these parameters are given by
\begin{eqnarray}
&&\frac{E}{A} - m_N = -16~{\rm MeV}, \quad n_0 = 0.16~{\rm fm}^{-3},\, \nonumber\\
&&K = 240 \pm 40~{\rm MeV}, \quad E_{\rm sym} = 31~{\rm MeV}\, ,\label{nmp}
\end{eqnarray}
Note that the compressibility $K$ has a relatively large uncertainty
compared to other nuclear matter properties.
Hence, we consider two different values of the compressibility as inputs, 
$K=215$ and $240$ MeV.
In Table~\ref{table:param_set1}, we summarize parameter sets used in this work.
These parameter sets are taken from Ref.~\cite{Shin:2018axs} except for the sets with $m_0 = 500$ 
with which binding energy and charge radius calculations do not converge.
In the nuclear structure studies~\cite{Shin:2018axs}, chiral invariant mass $m_0 = 700$ MeV is preferred.

\section{Application of The Extended Parity Doublet Model to Heavy Ion Collisions}
\label{sec:result}

As an application of EPDM to heavy ion collisions, 
we consider $^{197}$Au+$^{197}$Au collisions with our new transport code DJBUU.
In this work, we focus on the time evolution of the effective masses 
and anisotropic collective flow.

\subsection{Time Evolution of Effective Masses}

\begin{figure}
\centering
\subfigure[\, $K=215$ MeV]{\label{fig:a}\includegraphics[height=78mm]{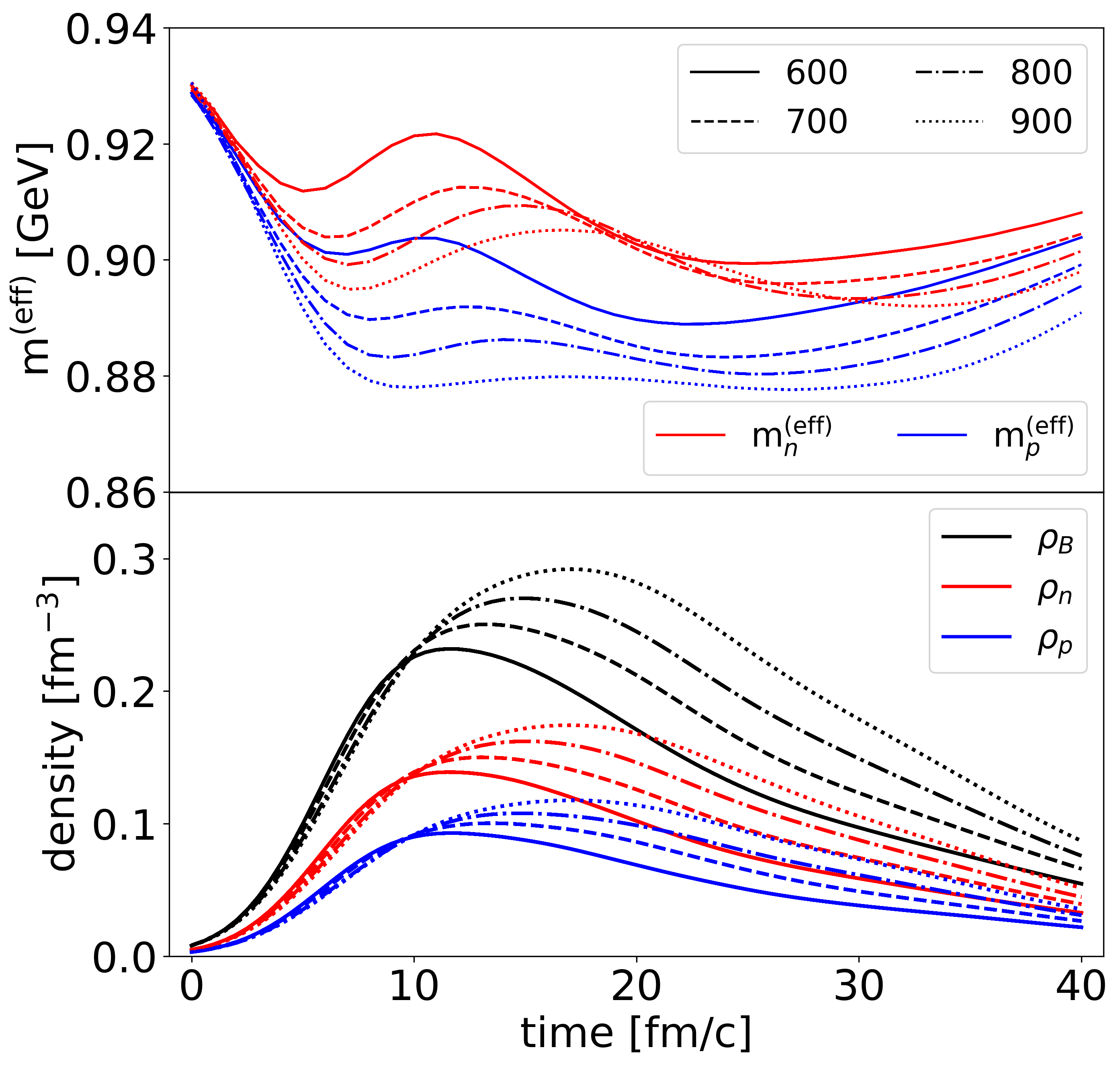}}\\
\subfigure[\, $K=240$ MeV]{\label{fig:b}\includegraphics[height=78mm]{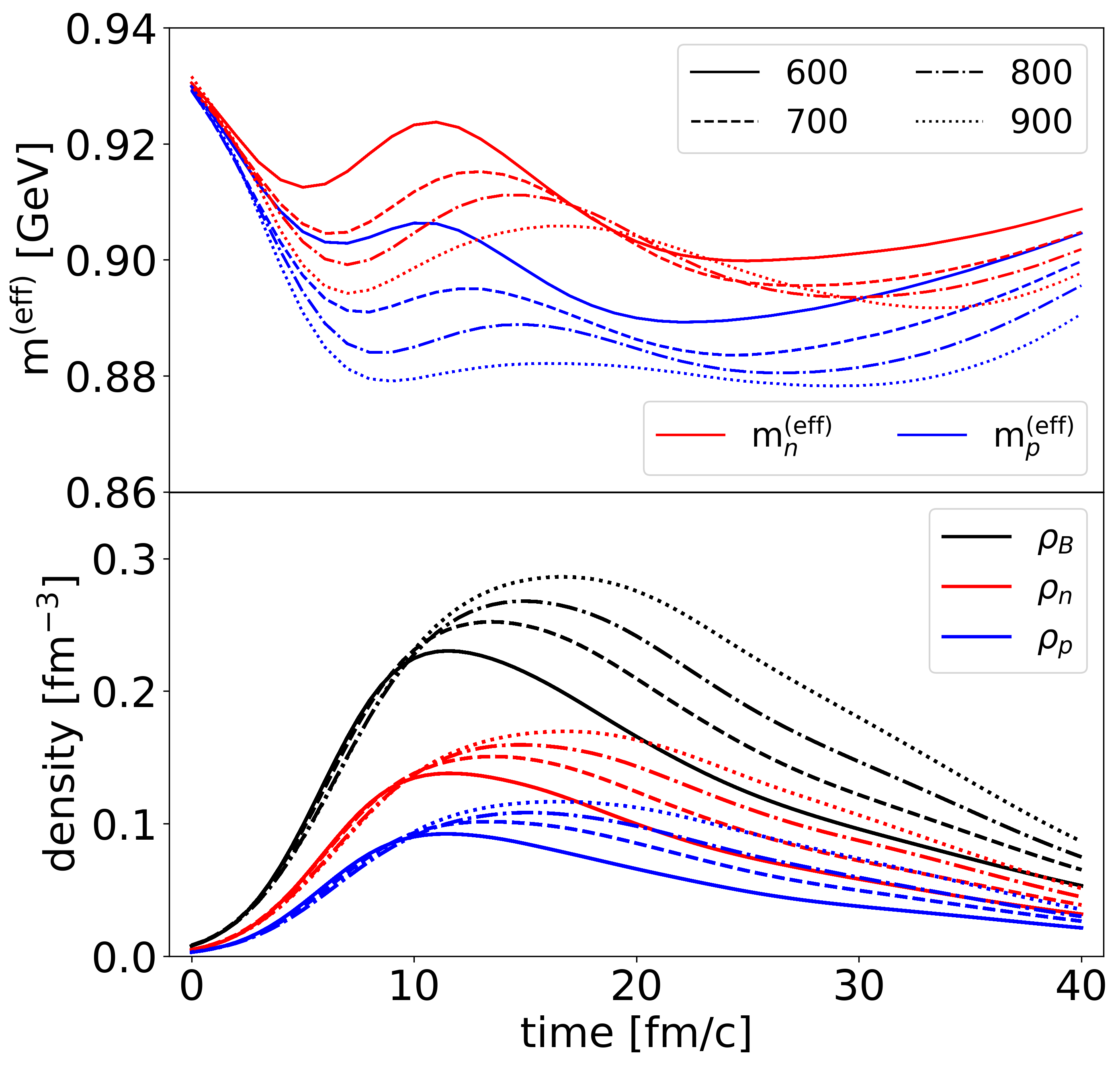}}
\caption{Time evolution of neutron and proton effective masses and densities at the center in 
$^{197}$Au+$^{197}$Au head-on collisions with $E_{\rm beam}= 400A$ MeV and
different compressibilities $K=215$ and 240 MeV.
The red color indicates physical quantities of neutron while blue is for protons. 
The black color is representing baryon (neutron + proton) quantities.
The solid, dashed, dash-dotted, and dotted lines are
$m_0=600$, $700$, $800$, and 900 MeV, respectively.}
\label{fig:mass_splitting1}
\end{figure}

The energies required to produce new particles in dense medium can be obtained from the dispersion relation:
\begin{eqnarray}
E_n &=& \sqrt{{m_N}^2 + {\bf k}^2} +g_\omega\bar{\omega} - g_\rho\bar{\rho}\, , \nonumber \\
E_p &=& \sqrt{{m_N}^2 + {\bf k}^2} +g_\omega\bar{\omega} + g_\rho\bar{\rho}\, ,
\end{eqnarray}
where $E_{n,p}$ are the energies of a neutron and a proton
and 
$m_N$ is the density-dependent nucleon mass $m_+$
defined in Eq.~(\ref{eq:mass}). 
As in Ref.~\cite{Takeda:2017mrm}, we define the effective nucleon masses  
as energies at ${\bf k}=0$ from the dispersion relation:
\begin{eqnarray}
m_n^{(\rm eff)} = m_N + g_\omega\bar{\omega} - g_\rho\bar{\rho}\, ,\nonumber \\
m_p^{(\rm eff)} = m_N + g_\omega\bar{\omega} + g_\rho\bar{\rho}\, .
\label{eq:meff}
\end{eqnarray}
As in other mean field models, 
there are significant effective mass splitting between protons and neutrons as
the isospin density increases.

The figures in Fig.~\ref{fig:mass_splitting1},
we summarize the time evolution of effective masses at the central part in
$^{197}$Au+$^{197}$Au head-on collision at 400$A$ MeV.
From Eq.~(\ref{eq:meff}), one can see that 
the exchange of isospin-dependent $\rho$ mesons
causes mass splitting between protons and neutrons.

\begin{figure}
\centering
\includegraphics[width=75mm]{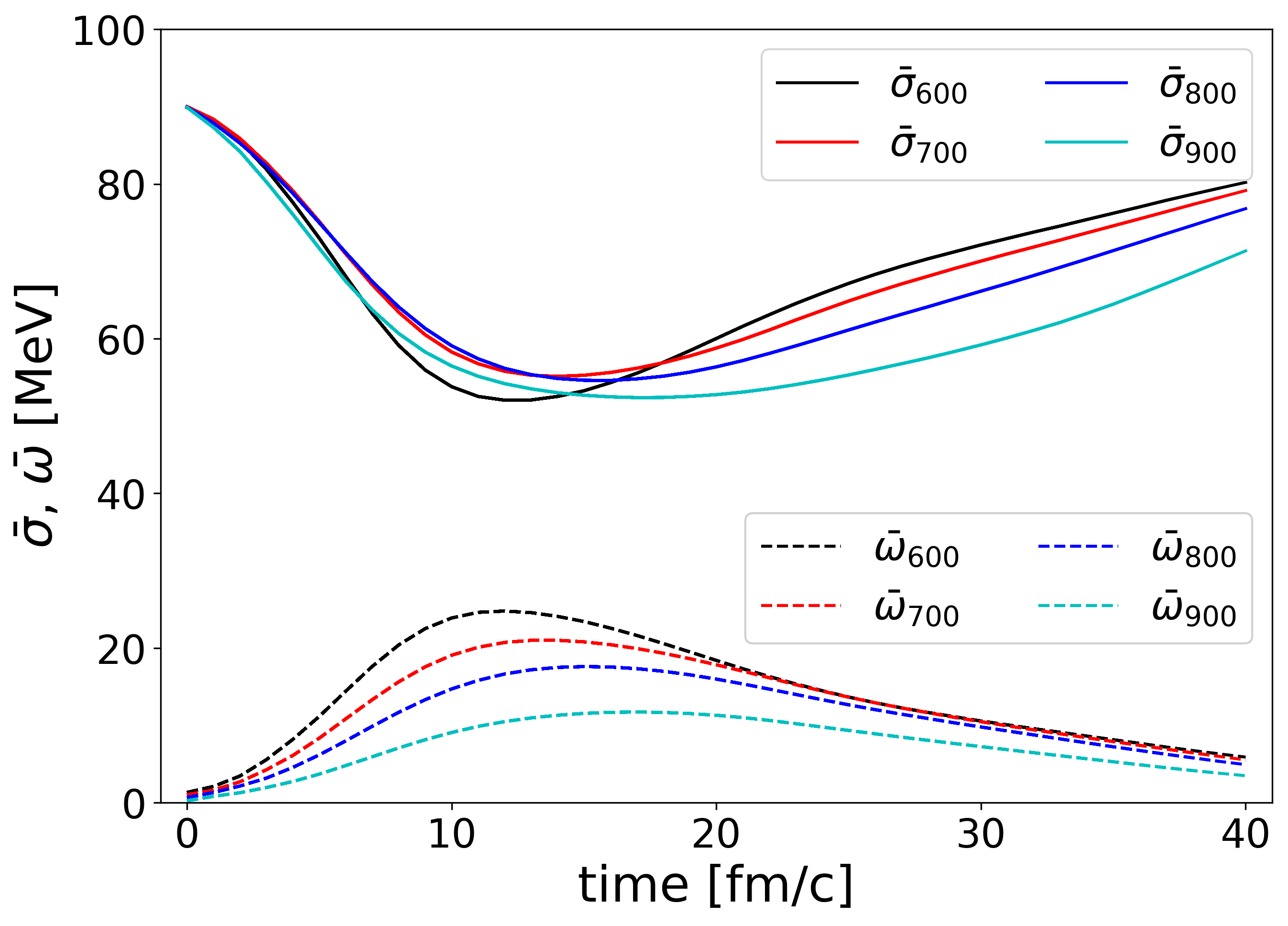}
\caption{Time evolution of $\bar{\sigma}$ and $\bar{\omega}$ for $^{197}$Au+$^{197}$Au central collision
at center position of center-of-mass frame.}
\label{fig:sigma_omega}
\end{figure}

The maximum value of the splitting increases as $m_0$ increases for both compressibilities.
But the maximum values of the splitting barely depend on the compressibility for a given $m_0$.
Maximum density increases as $m_0$ increases and lies in the range 
$1.44 < \rho_{max} / \rho_0 < 2.0$.
The $1\,\sigma$ statistical uncertainties in our calculations are
rather small and thus not shown in the figure. 
For instance, with $m_0$ = 700 MeV and $K=240$ MeV
the maximum density is calculated to be $0.2523 \pm 0.0032$ fm$^{-3}$.
One clear trend is that the maximum density increases as the chiral invariant mass $m_0$
increases. This behavior can be explained in terms of the behaviors of the $\sigma$ field and
the $\omega$ field.  In Fig.~\ref{fig:sigma_omega}, 
the expectation values of $\sigma$ and $\omega$ meson fields are summarized.
One can see that $\omega$ mean field decreases faster with the increasing
$m_0$ than the $\sigma$ mean field.
As $\omega$ provides repulsion and $\sigma$ provides attraction,
larger value of $m_0$ naturally results in the larger value of the nucleon density. 

If one can measure or estimate the maximum densities in HICs,
the value of the chiral invariant mass $m_0$ could be narrowed down.

\subsection{Anisotropic Collective Flow}

The heavy-ion collisions with a finite impact parameter develop
an anisotropic collective flow in momentum distribution.
Since the flow depends on the mean fields, collisions, blocking, etc.,
it can provide valuable information on dense medium.
In general, the flow can be quantified in terms of the Fourier expansion
of the momentum density in the azimuthal angle $\phi$~\cite{Ollitrault:1992bk}:
\be
{dN\over dy d^2 p_t} \propto 1 + 2\sum_{n=1}^\infty v_n(y,p_t)\cos(n(\phi-\psi_n))\, ,
\ee
where $\psi_n$ is the event plane angle for the $n$-th harmonics. 
The first two flow coefficients
$v_1$ and $v_2$ are often referred to as the direct and elliptic flows, respectively.
The flow coefficients $v_n(y,p_t)$ 
are the functions of rapidity $y$ and transverse momentum $p_t = \sqrt{p_x^2+p_y^2}$.
Here we focus on the directed flow defined as $v_1 = \left<p_x/p_t\right>$ 
for the particles with positive rapidity. Note that one can always set $\psi_1 = 0$ by
re-orienting the system.

\begin{figure}
\centering
\includegraphics[width=85mm]{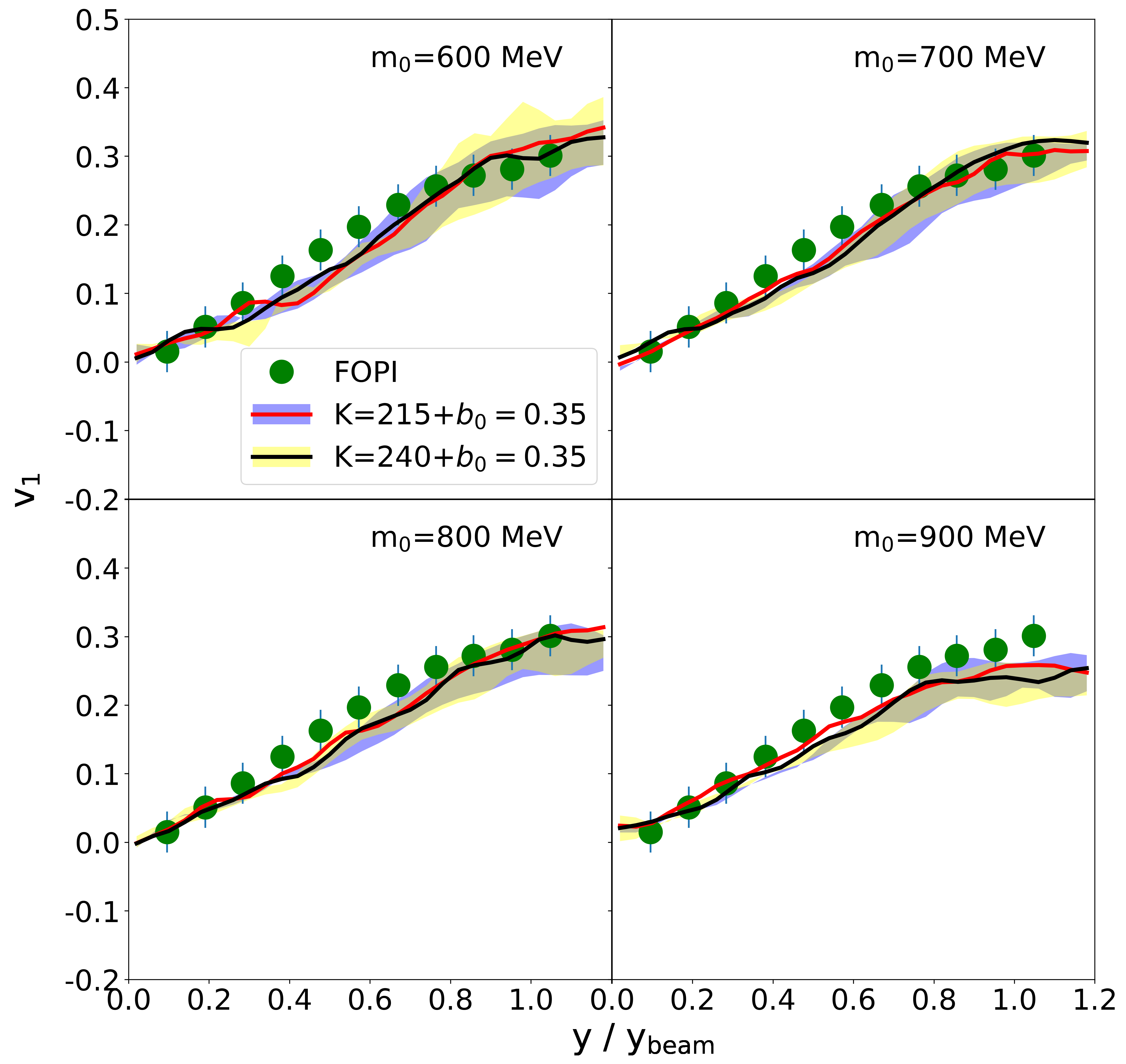}
\caption{Proton directed flow as a function of reduced rapidity 
for $^{197}$Au+$^{197}$Au collisions with $0.25< b_0 < 0.45$ at $E_{\rm beam}=400 A$ MeV. 
Two values of compressibility, $K=215$ MeV (purple shaded area)
and $K=240$ MeV (yellow shaded are), are considered.
Upper and lower limits of each shaded area correspond to the upper and lower limits of impact parameter $b_0$,
and the solid line correspond to the mean value $b_0=0.35$.
FOPI data are taken from Ref.~\cite{Xie:2014uia}.
}
\label{fig:v1_flow}
\end{figure}

The directed flow $v_1$ of protons
as a function of reduced rapidity is shown in Fig.~\ref{fig:v1_flow}.
The results shown are for the
$^{197}$Au+$^{197}$Au collisions at $E_{\rm beam} = 400 A$ MeV.
To match the FOPI cuts \cite{Reisdorf:2006ie,FOPI:2011aa},
we define two scaled parameters.
The scaled impact parameter is defined as $b_0 = b/b_{max}$
where $b_{max} = 1.15\times (A^{1/3}_P + A^{1/3}_T)$.
The scaled transverse velocity is defined as $u_{t0} = u_t/u_p$
where $u_t$ is the transverse component of the 4-velocity
of a particle and $u_p$ is the beam direction component of the 4-velocity
of the beam. The cuts we impose are $0.25<b_0<0.45$ and $u_{t0}>0.4$.

In Fig.~\ref{fig:v1_flow}, one can see
that the proton directed flows with  $m_0=600, 700$ and 800 MeV 
are all roughly consistent with experiments and there is not much
sensitivity to the compressibility.
One may say that the highest chiral invariant mass tested, $m_0=900$ MeV,
is disfavored because it deviates from the data at higher rapidities.
This can be again explained by the weaker $\omega$ field which would not provide
enough repulsion.
However, the deviation is not significant enough for a firm conclusion.

\begin{figure}
\centering
\includegraphics[width=70mm, height=60mm]{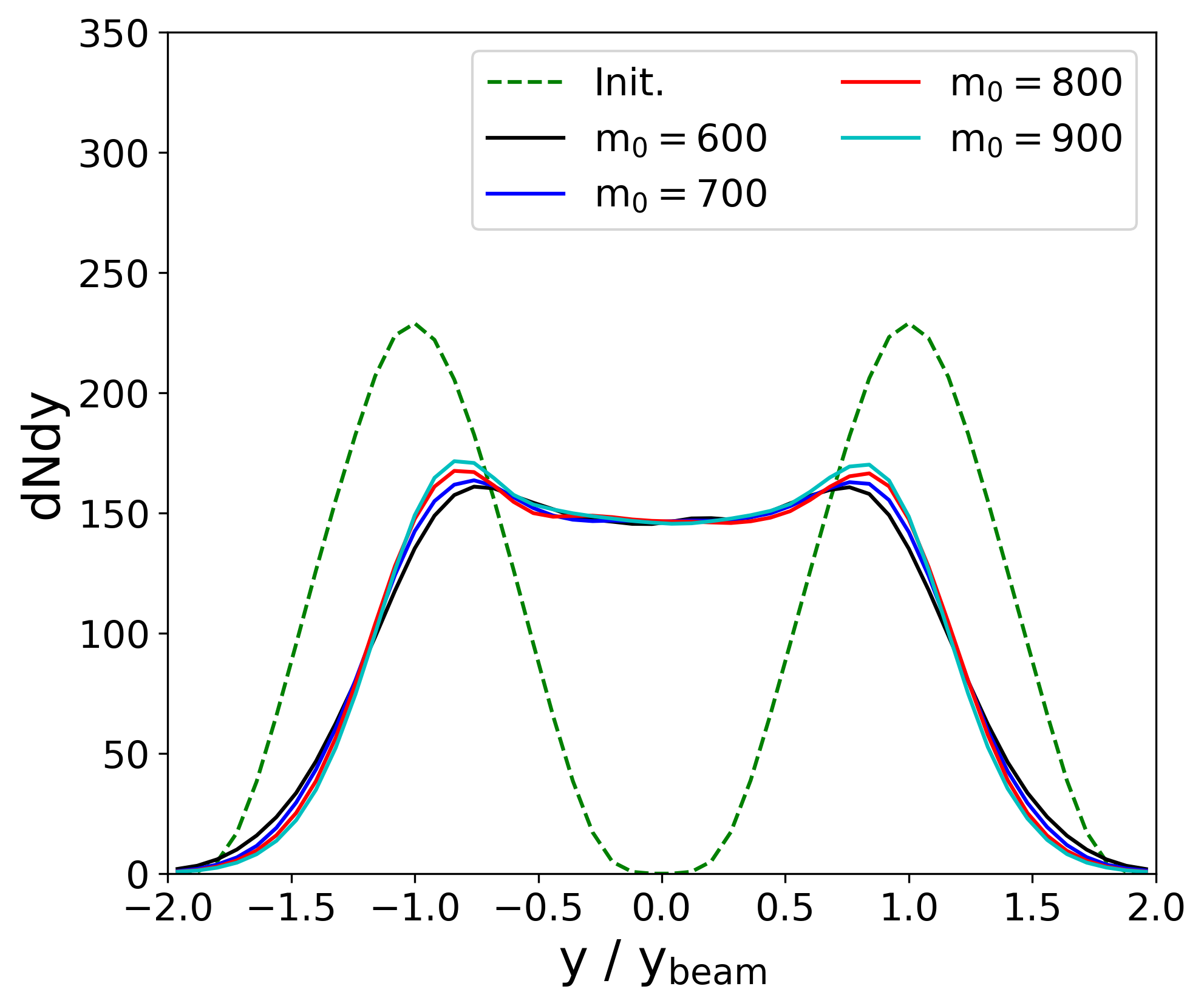}
\caption{Nucleon rapidity distributions at the beam energy of 400$A$ MeV at $b_0=0.35$ fm with $K=215$ MeV.}
\label{fig:dNdy}
\end{figure}

In Fig.~\ref{fig:dNdy}, the nucleon rapidity distributions of two nuclei
at  the initial time (dashed line) 
and the final time (solid lines) for four different chiral invariant masses are plotted. 
In this figure, only $K=215$ MeV is shown. Setting $K=240$ yields similar results.
The rapidity distribution along the beam axis reflects 
the nucleon stopping effects in HICs.
The initial distributions have peaks at $y/y_{\rm beam} = \pm 1$
because particles are distributed around the beam rapidities
at the initial time.
The stopping is largely insensitive to the value of the chiral invariant mass.

\section{Summary and Conclusions}
\label{sec:conclusion}

In this work, we have studied low-energy heavy ion collisions and infinite dense matter 
using DJBUU 
which is a new transport code of relativistic Boltzmann-Uehling-Uhlenbeck type. 
In order to test the validity of DJBUU, we compared our results with those
reported in the transport code comparison project studies.
We found that our results are consistent with the TCCP results, such as
nuclei stability, time evolution of density in Au+Au collisions, Pauli blocking and collisions, 
rapidity distribution, and collision itself in box calculations.

After confirming the validity of DJBUU, we implemented
the extended parity doublet model in DJBUU for the heavy ion collision simulations.
For the time evolution of effective masses in the medium,
we simulated central $^{197}$Au + $^{197}$Au collisions
at $E_{\rm beam} = 400A$ MeV for four different values of $m_0$.
In general, the mass splitting between protons and neutrons are found to increase
as the chiral invariant mass increases.
We also found that the results are not so sensitive to the compressibility.
The proton directed flow and rapidity distribution have been studied and compared with
the experimental result of FOPI.
We found that $m_0=600, 700, 800$ and  900 MeV give similar results as far as directed flow is concerned,
even though there are some deviations at the large rapidity region for $m_0 = 900$ MeV.

In the future, other nuclear models, such as 
KIDS~\cite{Papakonstantinou:2016zpe}, will be tested with DJBUU, and
our numerical calculations will be compared with the results 
from future rare isotope experiments within a few hundreds $A$ MeV.

\section*{Acknowledgements}
MK and CHL were supported by  National Research Foundation of Korea (NRF) grants funded by the 
Korea government (Ministry of Science and ICT and Ministry of Education)  (No. 2016R1A5A1013277 and No. 2018R1D1A1B07048599).
S.J. is supported in part by the Natural Sciences and Engineering Research Council of Canada.
Y.M.K was supported by NRF grants funded by the Korea government (No. 2016R1A5A1013277 and No. 2019R1C1C1010571).
The work of Y.K. was supported by the Rare Isotope Science Project of Institute for Basic Science funded by Ministry of
Science and ICT and National Research Foundation of Korea (2013M7A1A1075764).
Y.K. acknowledges useful discussions with Masayasu Harada.

\end{document}